\pdfoutput=1
\documentclass[journal]{IEEEtran}
\usepackage{amsmath,graphicx,kotex,siunitx}
\usepackage{amssymb}
\usepackage{amsfonts}
\usepackage{pifont}

\usepackage[flushleft]{threeparttable}
\usepackage{setspace}       

\ifCLASSOPTIONcompsoc
    \usepackage[caption=false, font=normalsize, labelfont=sf, textfont=sf]{subfig}
\else
\usepackage[caption=false, font=footnotesize]{subfig}
\fi
\usepackage{microtype}
\usepackage{graphics}
\usepackage{booktabs}
\usepackage{multirow}
\usepackage{placeins}
\usepackage{nccmath}
\usepackage[ruled,vlined]{algorithm2e}
\usepackage{xcolor}
\usepackage{hyperref}
\usepackage[normalem]{ulem}

\newsavebox\CBox

\usepackage{cite}

\begin{document}
\title{Differentiable Artificial Reverberation}
\author{Sungho~Lee, Hyeong-Seok~Choi, and~Kyogu~Lee \thanks{
Manuscript received October 21, 2021; revised March 27, 2022 and June 10, 2022; accepted July 15, 2022. 
This work was supported by Institute of Information \& communications Technology Planning \& Evaluation (IITP) grant funded by the Korea government (MSIT) (No.2022-0-00641).
The associate editor coordinating the review of this manuscript and approving it for publication was Cecchi, Stefania \textit{(Corresponding Author: Kyogu Lee)}.

Sungho Lee is with the Music and Audio Research Group, Graduate School of Convergence Science and Technology, Seoul National University, Seoul, Republic of Korea (e-mail: sh-lee@snu.ac.kr).

Hyeong-Seok Choi was with the Music and Audio Research Group, Graduate School of Convergence Science and Technology, Seoul National University, Seoul, Republic of Korea. He is now with the Institute of New Media and Communications, Seoul National University, Seoul, Republic of Korea (e-mail: kekepa15@snu.ac.kr).

Kyogu Lee is with the Music and Audio Research Group, Graduate School of
Convergence Science and Technology, Seoul National University, Seoul,
South Korea, and also with the Advanced Institutes of Convergence Technology,
Suwon, South Korea (e-mail: kglee@snu.ac.kr).

}}

\markboth{Lee \MakeLowercase{\textit{et al.}}: Differentiable Artificial Reverberation}
{Lee \MakeLowercase{\textit{et al.}}: Differentiable Artificial Reverberation}

\maketitle

\begin{abstract}
Artificial reverberation (AR) models play a central role in various audio applications. 
Therefore, estimating the AR model parameters (ARPs) of a reference reverberation is a crucial task.
Although a few recent deep-learning-based approaches have shown promising performance, their non-end-to-end training scheme prevents them from fully exploiting the potential of deep neural networks.
This motivates the introduction of differentiable artificial reverberation (DAR) models, allowing loss gradients to be back-propagated end-to-end. However, implementing the AR models with their difference equations “as is" in the deep learning framework severely bottlenecks the training speed when executed with a parallel processor like GPU due to their infinite impulse response (IIR) components.
We tackle this problem by replacing the IIR filters with finite impulse response (FIR) approximations with the frequency-sampling method. 
Using this technique, we implement three DAR models---differentiable Filtered Velvet Noise (FVN), Advanced Filtered Velvet Noise (AFVN), and Delay Network (DN).
For each AR model, we train its ARP estimation networks for analysis-synthesis (RIR-to-ARP) and blind estimation (reverberant-speech-to-ARP) task in an end-to-end manner with its DAR model counterpart.
Experiment results show that the proposed method achieves consistent performance improvement over the non-end-to-end approaches in both objective metrics and subjective listening test results. Audio samples are available at \href{https://sh-lee97.github.io/DAR-samples/}{https://sh-lee97.github.io/DAR-samples/}.
\end{abstract}


\IEEEpeerreviewmaketitle

\section{Introduction} \label{intro}
    Reverberation is ubiquitous in a real acoustic environment. It provides the listeners psychoacoustic cues for spatial {characteristics}. Therefore, adding an appropriate reverberation to a dry audio is {desirable} for plausible listening \cite{bayless1957innovations, blesser2001an, traer2016reverbperception}. \emph{Artificial reverberation} (AR), {efficient digital filters that may be used to mimic real-world reverberation}, have been developed to achieve such auditory effects \cite{valimaki20505years,valimaki16more, dattorro1997effect} and applied to room acoustic enhancement \cite{Gardner1992ARM}, auditory scene generation \cite{sena2015sdn, agus2018minimally}, post-production \cite{sarroff20revmatch}, and many more.
    
    \newpage
    Nevertheless, estimating the AR model parameters (ARPs) that match the {reference} reverberation remains challenging. This is because the mapping from the ARPs to the reverberation is highly nontrivial. We refer to this task as \emph{analysis-synthesis} when the {reference} is a room impulse response (room IR, RIR). When the reference is indirectly provided with a reverberant signal, i.e., an RIR convolved with a dry signal, we refer to this task as \emph{blind estimation}. We limit the scope to reverberant speech signals in this paper, yet the framework can easily be extended to other types of signals, such as musical ones.

    ARP estimation is a classical problem, and various attempts have been made \cite{jot92fdnanalysissynthesis, coggin16fdnmatch, valimaki2017fvn, peters2012matching}. However, they are only applicable to a specific AR model (AR-model-dependent) and the analysis-synthesis task (task-dependent). This inflexibility is problematic since every application has its own suitable AR models and {reference} forms (e.g., RIR and reverberant speech).

    \begin{figure}[!t]
    \centering
    \subfloat[Training ARP estimation network with DAR models. \label{fig:framework-train}]{
        \includegraphics[width=.98\columnwidth]{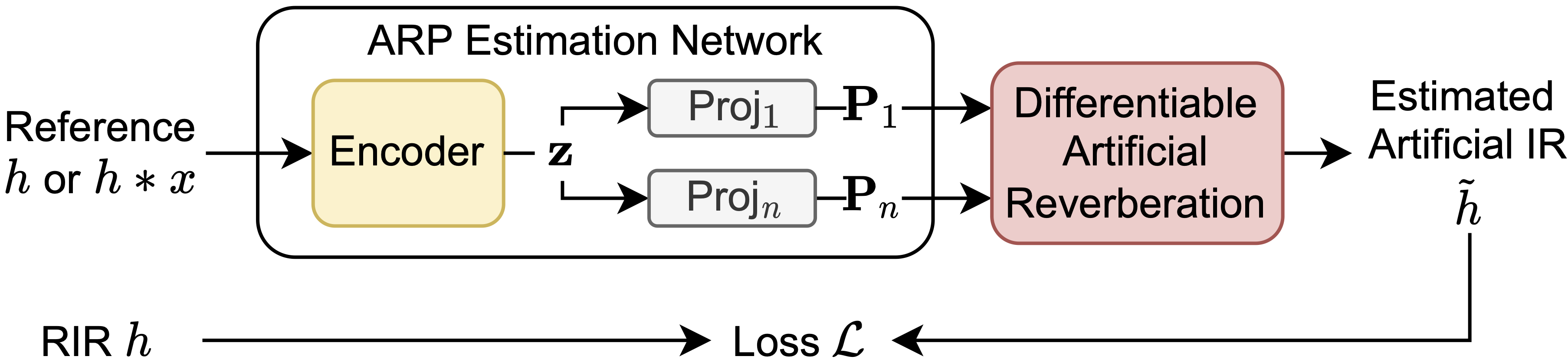}} \\
    \subfloat[Inference-time usage of the trained ARP estimation network. \label{fig:framework-test}]{
        \includegraphics[width=.98\columnwidth]{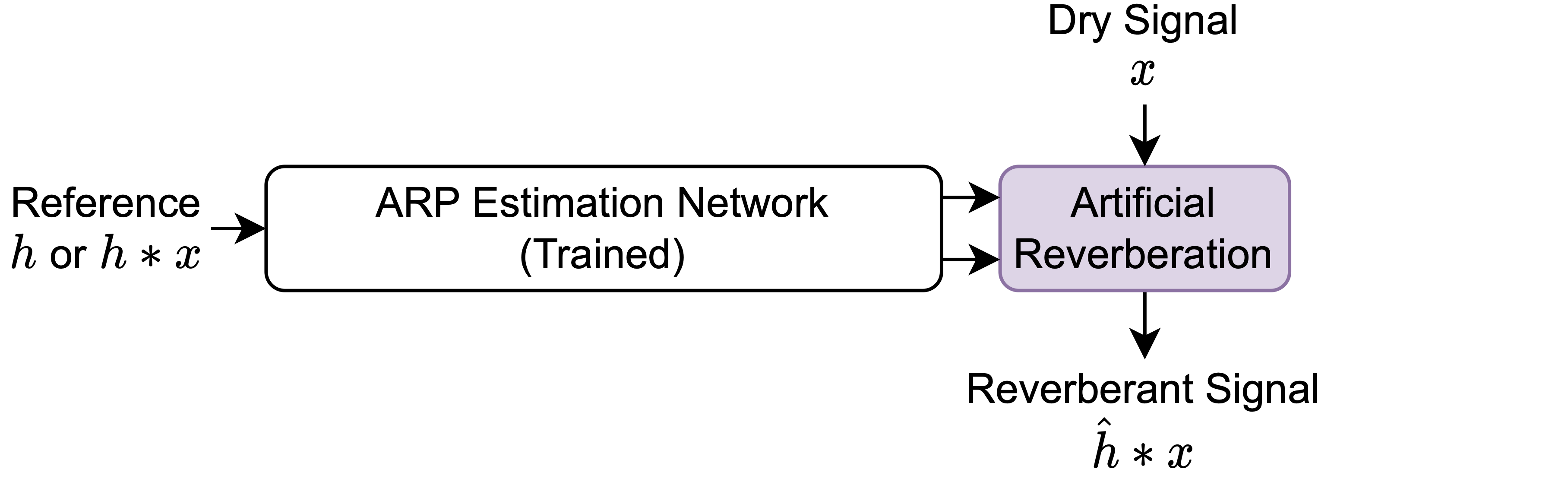}} \\
    \caption{The proposed ARP estimation framework. (a) With the DAR models, loss gradients $\partial \mathcal{L} / \partial \mathbf{P}_1, \cdots, \partial \mathcal{L} / \partial \mathbf{P}_n$ can be back-propagated through the DAR models. Hence, the ARP estimation network can be trained in an end-to-end manner. We can train the network to perform the analysis-synthesis (RIR $h$ input), blind estimation (reverberant speech $h*x$ input), or even both. The network has an AR-model-agnostic encoder so that using a different DAR model only requires changing the tiny projection layers $\text{Proj}_1, \cdots, \text{Proj}_n$. (b) Each DAR model generate an FIR approximation of its original AR model's IIR. After the training, estimated ARPs can be plugged in to the AR model which is highly efficient and real-time controllable in CPU.}
  \label{fig:framework} 
\end{figure}
    We can overcome such limitations with \emph{differentiable artificial reverberation} (DAR) models. Each DAR model takes ARPs as input and generates its corresponding AR model's IR (see Figure \ref{fig:framework-train}). After computing a loss between the generated IR and the reference RIR, they allow the loss gradients to be back-propagated through themselves. As a result, we can train a deep neural network (DNN) to match the reference RIR in an end-to-end manner. By leveraging the DNN's flexibility and expressiveness, we can train the network to perform any task with any AR model with minimal architecture change.
    
    \newpage
    Since the AR models are exactly or close to linear time-invariant (LTI), we can implement them differentiably “as is”  with their difference equations\cite{frank92iir, back91firiir, campolucci96iirmlp, kuznetsov2020diffiir}. 
    However, this approach has been impractical because the training speed is limited by the recurrent sample generation of their IIR filter components when executed on a parallel processor, such as a GPU.
    In this context, we aim to modify the vanilla DAR models to sidestep the aforementioned problem. Inspired by recent works \cite{neuralbiquads, nercessian2021lightweight}, we replace the IIR filters with finite impulse response (FIR) approximations using the frequency-sampling method \cite{rabiner70freqsamp}. This way, each DAR model becomes an FIR filter whose IR can be generated without any recurrent step. Using this technique, we present differentiable \emph{Filtered Velvet Noise} (FVN) \cite{valimaki2017fvn}, \emph{Advanced Filtered Velvet Noise} (AFVN), and \emph{Delay Network} (DN) \cite{puckette1989reverb, gerzon1971reverb}. Note that the AR-to-DAR conversion can be applied systematically to any other AR model with IIR components. 
    While the DAR models are not identical to their originals, the differences are little in most cases, allowing us to train the ARP estimation networks reliably with them.
    Furthermore, even when differences exist, the benefits of the end-to-end learning enabled by the DAR models outweigh them; we show that the proposed approach brings consistent performance improvement over non-end-to-end baselines where the difference is only in the training scheme. 
    After the training, we can revert to the original AR models for efficient computation (see Figure \ref{fig:framework-test}). 
    
\section{Related Works} \label{section:related-works}
    \subsection{Differentiable Digital Signal Processing}
    Digital signal processing (DSP) and deep learning are closely related. For example, a temporal convolutional network \cite{bai2018empirical} is a stack of dilated one-dimensional convolutional layers with nonlinear activation layers and it has been used for modeling dynamic range compression \cite{steinmetz2021efficient} and reverberation \cite{steinmetz2020diffmixconsole}. From the DSP viewpoint, it is FIR filter banks serially connected with memoryless nonlinearities, i.e., an extension of the classical Wiener-Hammerstein model \cite{BILLINGS198215}. 
    
    If we look for more explicit DSP-deep learning relationships, so-called “differentiable digital signal processing (DDSP)" approaches exist which aim to import DSP components into deep learning frameworks as they provide strong structural priors and interpretable representations. 
    For example, a harmonic-plus-noise model \cite{serra1990analysissynthesis} was implemented in such a manner; then, a DNN was trained to estimate its parameters to synthesize a monophonic signal and used for various applications, including controllable synthesis and timbre transfer \cite{engel2020ddsp}.
    
    In the case of LTI filters, it is well-known that both an FIR and IIR filter are differentiable and can be directly imported into the deep learning framework as a convolutional layer and recurrent layer, respectively \cite{frank92iir, back91firiir, campolucci96iirmlp, kuznetsov2020diffiir}. Since the IIR filter, e.g., parametric equalizer (PEQ), bottlenecks the training speed, the frequency-sampling method \cite{rabiner70freqsamp} was utilized \cite{neuralbiquads, nercessian2021lightweight}.  Upon these works, we investigate the reliability of this technique, present a general differentiable IIR filter based on state-variable filter (SVF) \cite{vafilter, Wishnick2014TimeVaryingFF}, and use it to implement the DAR models. 

    \subsection{ARP Estimation}
    \subsubsection{Analysis-synthesis}
    Various methods have been proposed to systematically and automatically estimate ARPs that mimic certain characteristics of a given RIR.
    Some works first extract perceptually relevant features from the RIR, then estimate the ARPs from them.
    For example, FDN parameters were derived from the RIR's energy decay relief (EDR) representation which encodes the frequency-dependent reverberation decay  \cite{jot92fdnanalysissynthesis}.
    Along with the FVN and AFVN, their parameter estimation algorithm based on linear prediction were proposed together \cite{valimaki2017fvn}.
    Another thread of work utilizes optimization.
    A genetic algorithm was proposed to find the FDN parameters that achieve the desired “fitness” to the reference RIR \cite{coggin16fdnmatch}. 
    Since all these methods assume availability of the RIR, they cannot be applied to the blind estimation task. 
    
    \subsubsection{Blind Estimation}
    End-users (e.g., audio engineers) often interact with the AR models provided as audio plug-ins. As a result, several attempts have been made to implement blind estimation algorithms for the plug-ins. For example, a preset recommendation method based on the Gaussian mixture model was proposed \cite{peters2012matching}. 
    More recently, DNNs were trained for the plug-in parameter estimation and preset recommendation \cite{sarroff20revmatch}. 

    While these methods have shown promise, there is still room for improvement.
    First, they generated the training data using the plug-ins to train their models. 
    However, this procedure may be time-consuming, and relying on data generated with a single plug-in may increase the generalization error. 
    Second, their training objectives could be suboptimal. When the task is defined as parameter regression \cite{sarroff20revmatch}, one must weigh each parameter's perceptual importance and apply it to the loss function design, which is a highly challenging task.
    When the task is defined as classification \cite{peters2012matching, sarroff20revmatch}, it could suffer from scalability problems.
    Finally, most plug-ins provide only a few parameters, and each of them controls multiple internal ARPs simultaneously.
    This might limit the performance due to the reduced degree of freedom. 
    
    We tackle these problems with the DAR models as follows.
    First, we can use a training objective that directly compares the estimated IR with ground-truth RIR. 
    Therefore, we can use any RIR for training and bypass the need to obtain the ground-truth ARPs.
    Second, we replace the parameter-matching loss with the end-to-end IR-matching loss and show that it improves the estimation performance by a large margin in both objective metrics and subjective listening test scores.
    Finally, we estimate the ARPs directly instead of relying on the plug-in parameters.

    \subsubsection{Two-Stage Approaches} \label{subsection:two-stage}
    Note that both analysis-synthesis and blind estimation tasks can be divided into two subtasks. First, we can obtain reverberation parameters, e.g., reverberation time, from a given reference (reference-to-reverberation parameters). We can directly compute the parameters from the RIR \cite{2008ISO3382_2} for the analysis-synthesis case. For the blind estimation, various methods have been proposed \cite{rama2003rtestimate, wen2008rtestimate, gamper2016rtestimatecnn, li2019freqrt_estimate}. Then, we can tune the ARPs to match the desired reverberation parameter values (reverberation parameters-to-ARPs) \cite{sebastian17revtimecontrol, karolina19imprevtimecontrol, chourdakis2017a}. This two-stage approach can be a possible alternative to our single-stage end-to-end estimation method. 
    However, perceptually different reverberations can have the same reverberation parameters, resulting in information loss.
    
    \subsection{Modeling Reverberation in the Deep Learning Framework}
    An alternative to the proposed DAR approach could be using existing deep learning modules to build another reverberation model. The most simplest approach could be using a high-order FIR directly as an RIR model \cite{engel2020ddsp}. While it is viable when the objective is to learn a single RIR, training a DNN to generate various raw RIRs is highly challenging and inefficient. 
    
    Instead, recent approaches either use a DNN as a reverberator and control them via conditioning methods \cite{steinmetz2020diffmixconsole, richard2021neural} or design a differentiable RIR generator and use a DNN as a parameter estimator \cite{steinmetz2021filtered}. Our method is closely related to the latter since the DAR model and the proposed ARP estimation network act in that way. However, the key difference is that our DAR models are direct imports of the AR models, which bring the following benefits.
    First, they are more directly interpretable and controllable, allowing end-users to further tune the parameters. Second, their structure can prevent unexpected artifacts. Finally, they use sparse IIR and FIR filters, enabling efficient real-time computation in CPU.
    
\section{Frequency-Sampling Method for Differentiable IIR Filters} \label{section:frequency-sampling}
    For any real IIR filter $H$, its frequency response $H(e^{j\omega})$ is continuous, $2\pi$-periodic, and conjugate symmetric. To obtain its FIR approximation, we \emph{frequency-sample} $H(e^{j\omega})$ at angular frequencies $\omega_k = 2\pi k/N $ where $k = 0, \cdots, \left\lfloor N / 2\right\rfloor$ \cite{rabiner70freqsamp}. We denote the frequency-sampled filter with subscript $N$, i.e., $H_N$.
    \begin{equation}
       H_N[k] = 
       H(e^{j\omega_k}).
    \end{equation} 
    Note that the frequency-sampling can also be defined on transfer functions, e.g., $H(z)$ with $z=e^{j\omega_k}$.
    
    Since each sampling is independent, $H_N$ can be generated simultaneously with a parallel processor like GPU. Furthermore, the order of frequency-sampling $(\cdot)_N$ and other basic arithmetic operations does not matter. Therefore, we can frequency-sample an IIR filter by combining frequency-sampled $m$-sample delays $(z^{-m})_N\in \mathbb{C}^{\left\lfloor N / 2 \right\rfloor + 1}$ as follows,
    \begin{equation}
        H_N[k] = \left(\frac{\sum_{m=0}^M \beta_m z^{-m}}{\sum_{m=0}^M \alpha_m z^{-m}}\right)_N[k] = \frac{\sum_{m=0}^M \beta_m (z^{-m})_N[k]}{\sum_{m=0}^M \alpha_m (z^{-m})_N[k]}. \label{diffiir}
    \end{equation} 
    Since $(z^{-m})_N[k]=e^{-j2\pi km/N}$, using the widely-used deep learning libraries, we obtain $(z^{-m})_N $ as a tensor $\texttt{z\_m}$ by\begin{subequations}
        \begin{align}
            \texttt{angle }&\texttt{= 2*pi*arange(N//2+1)/N}, \\
            \texttt{z\_m }&\texttt{= e**(-1j*angle*m)}.
        \end{align}
    \end{subequations}
    Then, we apply equation \eqref{diffiir} to $\texttt{z\_0}, \cdots, \texttt{z\_M}$ with basic tensor operations and obtain every sample of $H_N$ in parallel. A time-domain representation of $H_N$ is a length-$N$ FIR $h_N[n]$, inverse discrete Fourier transform (inverse DFT) of $H_N[k]$, i.e.,
    \begin{equation}
    h_N[n] = \frac{1}{N}\sum_{k=0}^{N-1} H_N[k] e^{j\omega_kn} (u[n] - u[n-N])
    \end{equation} 
    where $u[n]$ is a unit step function. 
    
    Therefore, using the frequency-sampling method, we can replace any IIR filter $H$ that causes the bottleneck with an FIR $h_N[n]$ whose convolution can be performed efficiently via Fast Fourier Transform (FFT). In this paper, the ``differentiable IIR'' filter refers to $H_N$, which is in fact an FIR filter.
    
    \subsection{Reliability of the Frequency-Sampling Method}\label{section:reliability}
\begin{figure}[t]
    \begin{center}
        \centerline{\includegraphics[width=\columnwidth]{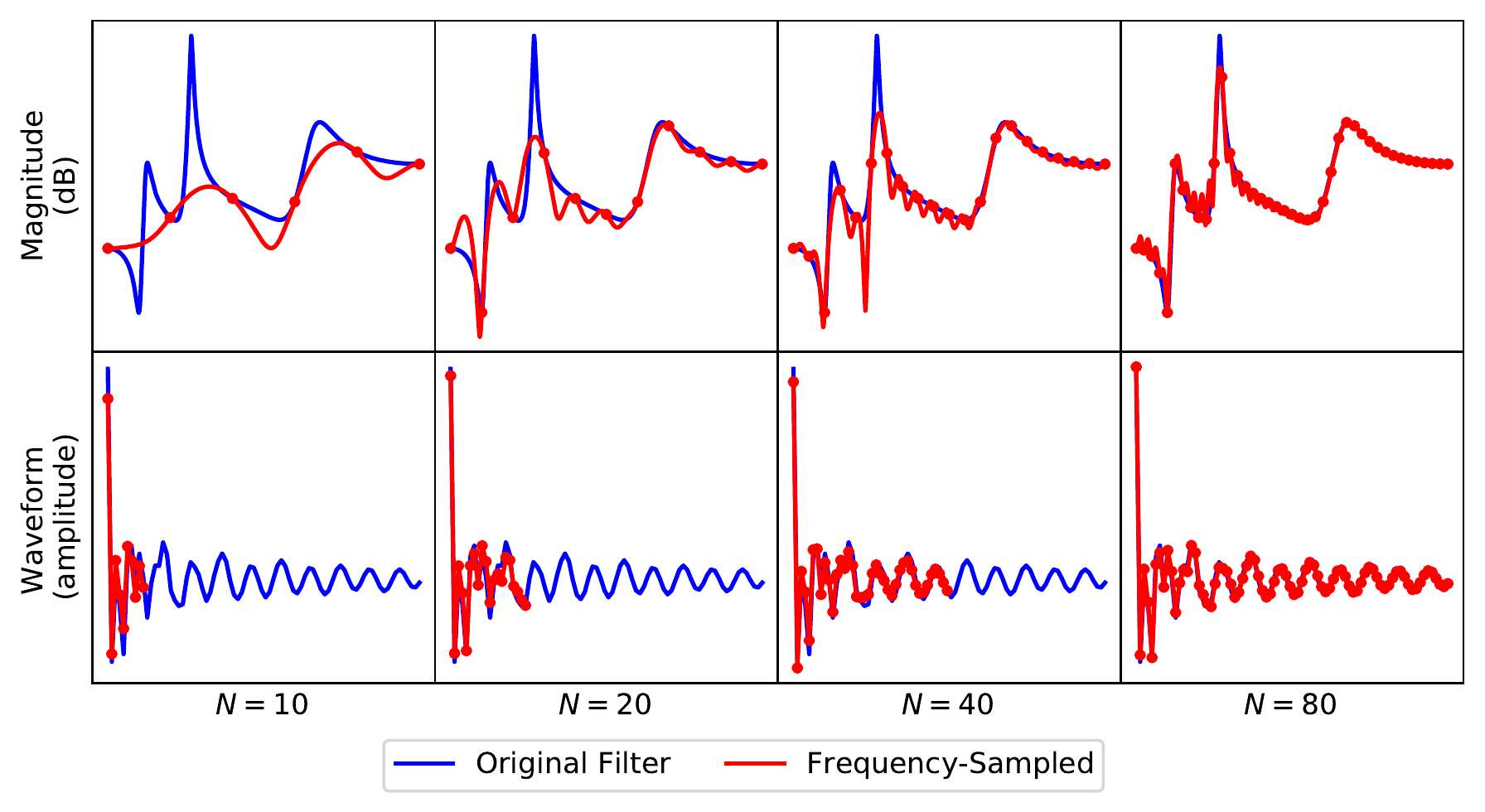}}
        \caption{The frequency-sampling method with a various number of sampling points $N$. 
        Blue curves represent its magnitude response $|H(e^{j\omega})|$ (top row) and IR $h[n]$ (bottom row). Red curves are the magnitude response and IR of the frequency-sampled filter $H_N$. 
        }
        \label{fig:frequency-sampling}
    \end{center}
\end{figure}

        Every IIR filter's frequency response has ``infinite bandwidth" (in the time domain) so that the frequency-sampling causes  \emph{time-aliasing} \cite{MDFT07}. 
        Specifically, the frequency-sampled IR $h_N[n]$ is sum of length-$N$ segments of the original IR $h[n]$.
        \begin{equation}
            h_N[n] = \sum_{m=0}^{\infty} h[mN + n] (u[n] - u[n-N]).
            \label{eq:time-alias}
        \end{equation}
        Since the energy of every stable IIR filter's IR decays over time, more frequency-sampling points $N$ give less time-aliasing, i.e., a closer approximation (see Figure \ref{fig:frequency-sampling}). 
        Therefore, we can use the differentiable filter $H_N$ in place of the original $H$ to match a {reference response $H_\text{Ref}$}. We elaborate this with following analytical results.
        \subsubsection{Time-aliasing Error} If $H$ has $M$ distinct poles where each of them $\nu_i \in \mathbb{C}$ has multiplicity $r_i \in \mathbb{N}$, the time-aliasing error asymptotically decreases as $N$ increases as follows, 
        \begin{equation}
            \|H-H_N\|_2 = \sum_{i=1}^{M} O(N^{r_i-1} |\nu_i|^{N}). \\
            \label{eq:fr-deviation}
        \end{equation}
        \subsubsection{Loss Error} 
        The triangle inequality indicates that the loss error is bounded to the time-aliasing error. 
        \begin{equation}
            \left|\|{H_\text{Ref}}-H\|_2^2 - \|{H_\text{Ref}}-H_N\|_2^2\right| \leq \|H-H_N\|_2^2.
            \label{eq:loss-bound}
        \end{equation}
        
        \subsubsection{Loss Gradient Error} The loss gradient error has similar asymptotic behavior to the time-aliasing error as follows,
        \begingroup
        \setlength{\thinmuskip}{1mu}
        \setlength{\medmuskip}{2mu}
        \setlength{\thickmuskip}{3mu}
        \begin{equation}
            \left|\frac{\partial}{\partial p}\|{H_\text{Ref}}-H\|_2^2 -\frac{\partial}{\partial p}\|{H_\text{Ref}}-H_N\|_2^2 \right| \\ 
            = \sum_{i=1}^{M} O(N^{3r_i-1}|\nu_i|^{N})
            \label{eq:deviation-loss}
        \end{equation}
        \endgroup
        where $p$ is a parameter of $H$. 
        See Appendix \ref{appendix:freq-samp} for the proofs.

    \subsection{Differentiable Dense IIR Filter with State-variable Filters}\label{subsection:differentiable-iir}
        Since arbitrary IIR filter can be expressed as serially cascaded biquads (second-order IIR filters), we can obtain a differentiable IIR filter $\Tilde{H}$ as a product of frequency-sampled biquads $(H_i)_N$.
        \begin{equation}
            \Tilde{H}[k] = \prod_i (H_i)_N[k] = \prod_i \left(\frac{\sum_{m=0}^2 \beta_{i, m} z^{-m}}{\sum_{m=0}^2 \alpha_{i, m} z^{-m}}\right)_N[k].    
        \end{equation}
        Moreover, we use the \emph{state variable filter} (SVF) parameters $f_i, R_i$, $m^\text{LP}_i$, $m^\text{BP}_i$, and $m^\text{HP}_i$ to express each biquad as follows,\begin{subequations}
            \begin{align}
                    \beta_{i, 0} &= f^2_i m^{\text{LP}}_i+f_i m^{\text{BP}}_i+m^{\text{HP}}_i, \label{svf_to_biquad_1} \\
                    \beta_{i, 1} &= 2f^2_i m^{\text{LP}}_i - 2m^{\text{HP}}_i, \\
                    \beta_{i, 2} &= f^2_i m^{\text{LP}}_i-f_i m^{\text{BP}}_i+m^{\text{HP}}_i, \\
                    \alpha_{i, 0} &= f^2_i + 2R_if_i + 1, \\
                    \alpha_{i, 1} &= 2f^2_i-2, \\
                    \alpha_{i, 2} &= f^2_i - 2R_if_i + 1. \label{svf_to_biquad_2} 
            \end{align}
        \end{subequations}

        We prefer to use the SVF parameters rather than the biquad coefficients due to the following reasons.
        \begin{itemize}
            \item \textit{Interpretability.} Each SVF has lowpass $H_i^\text{LP}$, bandpass $H_i^\text{BP}$, and highpass filter $H_i^\text{HP}$. They share a resonance $R_i$ and cutoff parameter $f_i = \tan(\pi \omega_i / \omega_s)$ where $\omega_i$ is their cutoff frequency and $\omega_s$ is the sampling rate. The filter outputs are multiplied with gains $m^\text{LP}_i$, $m^\text{BP}_i$, and $m^\text{HP}_i$ and then summed. 
            This interpretability comes without any generality loss since an SVF can express any biquad \cite{Wishnick2014TimeVaryingFF}. 
            \item \textit{Simple Activation Functions.} The SVF parameters have simpler stability conditions, $R_i>0$ and $ f_i >0 $, than the biquad coefficients \cite{kuznetsov2020diffiir}, which leads to simpler activation function design. Refer to Section \ref{subsection:activ-init} for details.
            \item \textit{Better Performance.} We found that our ARP estimation networks performed better when they estimate the SVF parameters rather than the biquad coefficients. Refer to Appendix \ref{subsection:coloration-comparison} for the comparison results.
        \end{itemize} 
        
        From now on, we denote each SVF-parameterized biquad with a superscript $H_i^\text{SVF}$ and call it simply ``SVF''. 
    
    \subsection{Differentiable Parametric Equalizer} \label{subsection:differentiable-peq}
        A low-shelving $H^\text{LS}$, peaking $H^\text{Peak}$, and high-shelving filter $H^\text{HS}$ are widely used audio filters. Each of them can be obtained using the SVF $H^\text{SVF}(f, R, m^\text{LP}, m^\text{BP}, m^\text{HP})$ as follows \cite{vafilter},
        \begin{subequations}
            \begin{align}
                H^\text{LS}(f, R, G) &= H^\text{SVF}(f, R, G, 2R\sqrt{G}, 1), \\
                H^\text{Peak}(f, R, G) &= H^\text{SVF}(f, R, 1, 2RG, 1), \\
                H^\text{HS}(f, R, G) &= H^\text{SVF}(f, R, 1, 2R\sqrt{G}, G). 
            \end{align}
        \end{subequations}
        Here, $G$ is a new parameter that gives $20\log_{10}(G)\,\si{dB}$ gain. A parametric equalizer (PEQ) is serial composition of such filters.
        Following the recent works \cite{neuralbiquads, nercessian2021lightweight}, we use one low-shelving, one high-shelving, and $K-2$ peaking filters.
        \begin{equation}
            H^\text{PEQ}(z) = H^\text{LS}(z) H^\text{HS}(z) \prod_{i=1}^{K-2} H^\text{Peak}_i(z).
        \end{equation}
        Same as the IIR filter case, frequency-sampling the components and multiplying them results in a differentiable PEQ $H^\text{PEQ}_N$.

\section{Differentiable Artificial Reverberation} \label{section:differentiable-artificial-reverberation}
    In this section, we derive differentiable FVN, {AFVN}, and DN. We first briefly review the original AR models. Then, we obtain their DAR counterparts by frequency-sampling the IIR components, replacing them with FIRs. Also, we slightly modify the original models for efficient training, plausible reverberation, and better overall estimation performance.
    \subsection{Differentiable Filtered Velvet Noise}\label{subsection:differentiable-filtered-velvet-noise}
        \subsubsection{Filtered Velvet Noise} 
\begin{figure}[t]
    \begin{center}
        \centerline{\includegraphics[width=.91\columnwidth]{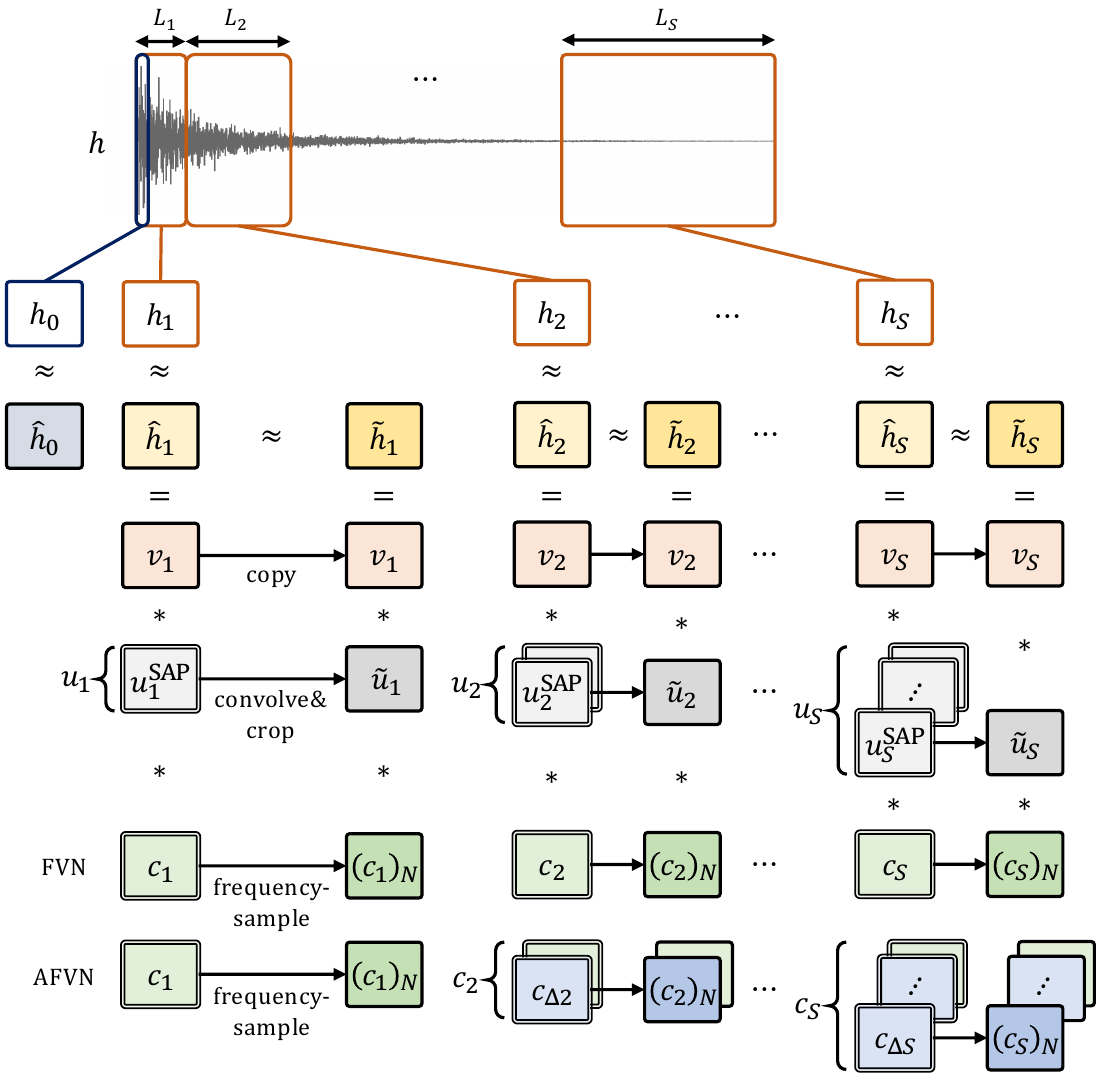}}
    \caption{Modified FVN, AVFN's RIR approximation, and their differentiable implementation strategy. FVN and AVFN divide a target RIR $h[n]$ into segments $h_i[n]$ and approximate each segment with a velvet noise $v_i$ filtered with an allpass filter $U_i$ and a coloration filter $C_i$. We obtain the differentiable FVN and AVFN by converting the IIR filters $U_i$ and $C_i$ into FIR filters $\Tilde{u}_i$ and $(c_i)_N$. Each IIR filter is emphasized with a double border box.
        }\label{fig:fvn}
    \end{center}
\end{figure}

            One can divide an RIR $h[n]$ into $S$ segments (see Figure \ref{fig:fvn}). As reverberation {can be regarded as} mostly stochastic, we can model each length-$L_i$ segment $h_i[n]$ with a source noise signal $s_i[n]$ ``colored with'' a filter $C_i$. When we use \emph{velvet noise} for each source, this source-filter model becomes \emph{Filtered Velvet Noise} (FVN) \cite{valimaki2013velvet, jarvelanen07velvet, valimaki2017fvn}. 
            The velvet noise $v_i[n]$ is sparse; in every length-$T_i$ interval, it contains a single nonzero sample $\pm 1$ with random sign/position such that its time-domain convolution is highly efficient. Moreover, an allpass filter $U_i$ is introduced to smooth each source and make plausible sound. 
            Finally, an deterministic (bypass) FIR $\hat{h}_0[n]$ models the direct arrival and early reflection. Then, an IR of FVN $\hat{h}[n]$ becomes
            \begin{equation}
                \hat{h}[n] = \hat{h}_0[n] + \sum_{i = 1}^{S} \underbrace{\overbrace{v_i[n - d_{i}] *u_i[n]}^{s_i[n-d_i]}*\;c_i[n]}_{\hat{h}_i[n - d_{i}]}
                \label{eq:fvn}
            \end{equation}
            where each $d_i$ is a delay for the segment alignment ($d_1 = 0$). In practice, each allpass filter $U_i$ is composed with Schroeder allpass filters (SAPs), $U_{i, j}^\text{SAP}(z) = (1+\gamma_{i, j} z^{-\tau_{i, j}})/(\gamma_{i, j} + z^{-\tau_{i, j}})$, where $\tau_{i, j}$ is a delay-line length and $\gamma_{i, j}$ is a feed-forward/back gain) for efficiency \cite{schroeder61colorless}. A low-order dense IIR filter is used as each coloration filter $C_i$.

        \subsubsection{Differentiable Implementation}
            We modify each FVN component to obtain differentiable FVN.
            \begin{itemize}
                \item \emph{Velvet filters}. To batch the IR segments, we use the same segment lengths $L_1, \cdots, L_S$ and average pulse distance values $T_1, \cdots, T_S$ for every {reference} RIR.
                \item \emph{Allpass filters}. We tune and fix the allpass filter parameters jointly with the velvet filter parameters to avoid perceptual artifacts, e.g., discontinuous and rough sound, while being computationally efficient. 
                Since we fixed the filter, instead of the frequency-sampling method, we simply crop its IR to obtain an FIR $\Tilde{u}_i[n]$. 
                \item \emph{Coloration filters}. We model $C_i$ with $K$ serial SVFs, i.e., $C_i(z)=\prod_{k=1}^{K} C_{i,k}^\text{SVF}(z)$. We frequency-sample each filter $C_i(z)$ and convert it into a length-$N$ FIR $(c_i)_N[n]$ so that we can train our network to estimate its parameters.
                
            \end{itemize}
            Therefore, IR of the differentiable FVN $\Tilde{h}[n]$ becomes
            \begin{equation}
                \Tilde{h}[n] = \hat{h}_0[n] + \text{overlap-add}\{\underbrace{\overbrace{v_i[n] * \Tilde{u}_i[n]}^{\Tilde{s}_i[n]} * (c_i)_N[n]}_{\Tilde{h}_i[n]}\}.
            \end{equation}
            Since every component $v_i[n]$, $\Tilde{u}_i[n]$, and $(c_i)_N[n]$ is an FIR, we can compute each segment $\Tilde{h}_i[n] = v_i[n] * \Tilde{u}_i[n]*(c_i)_N[n]$ efficiently by convolving the FIRs in the frequency domain, i.e., zero-padding followed by FFT, multiplication, then IFFT. Finally, we overlap-add all the segments and add the deterministic FIR $\hat{h}_0[n]$ to obtain the full IR $\Tilde{h}[n]$. The overlap-add can be implemented with a transposed convolution layer.
            \begin{figure}[!t]
    \centering
    \subfloat[Approximation error of differentiable FVN in EDR. \label{fig:approx-error-fvn}]{
        \includegraphics[width=.98\columnwidth]{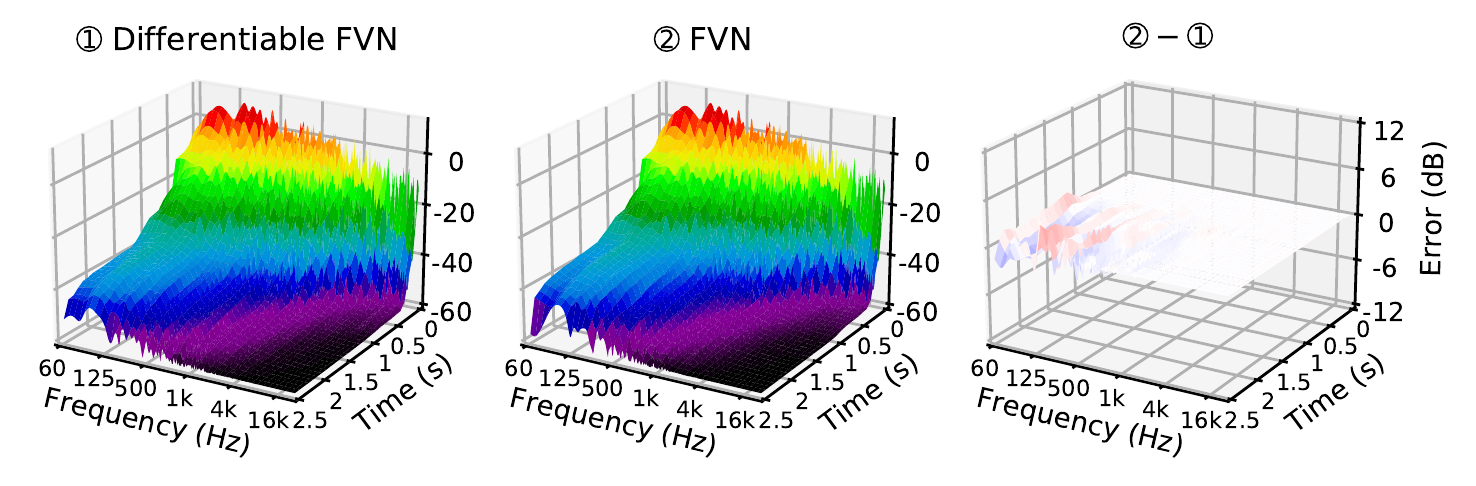}} \\
    \subfloat[Approximation error of differentiable AFVN in EDR.  \label{fig:approx-error-afvn}]{
        \includegraphics[width=.98\columnwidth]{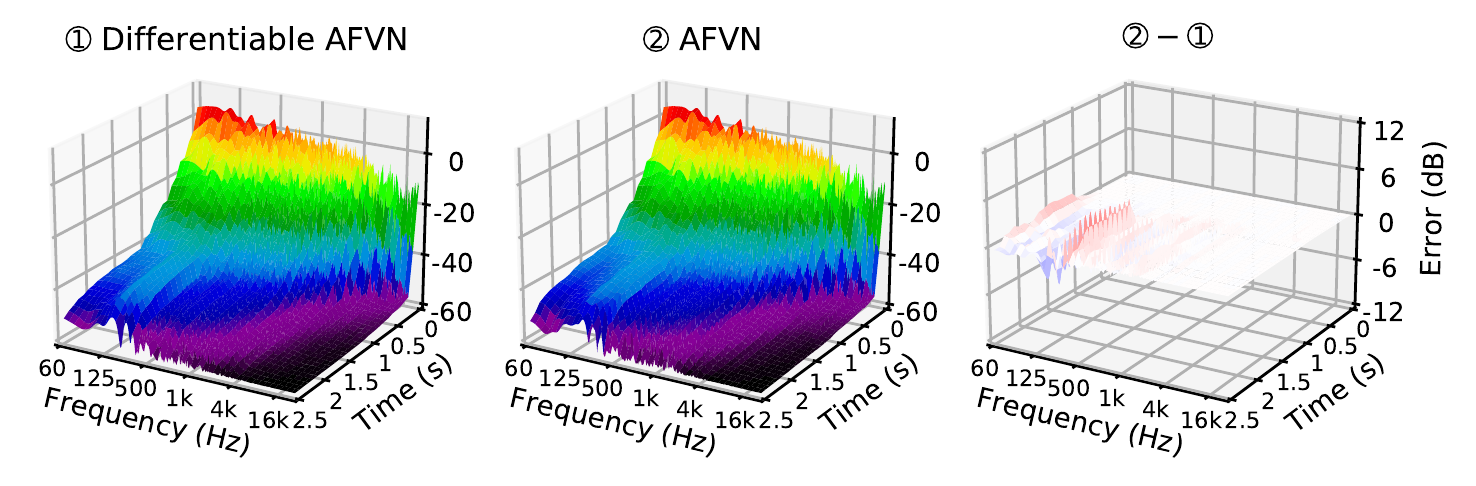}} \\
    \caption{EDR of IRs generated by differentiable FVN, AFVN, their original AR counterparts, and EDR errors introduced during the AR-to-DAR conversion process. Because the errors are small and mostly in the low-energy region, they are imperceptibly small (note that the $z$ axis scale is different for the EDR plots and the EDR error plots). We obtained the FVN/AFVN parameters with the trained blind estimation networks discussed below.} 
  \label{fig:approx-error-velvet} 
\end{figure}        
            
            While the modifications for the differentiable FVN introduce approximation error, it is negligible in practice. We explain this with the EDR $H^\text{EDR}$, a reverse cumulative sum of an IR's short-time Fourier transform (STFT) energy \cite{jot92fdnanalysissynthesis}.
            \begin{equation}
                H^\text{EDR}[k, n] = \sum_{m=n}^{\infty}|H^\text{STFT}[k, m]|^2.
            \end{equation}
            By comparing the FVN's EDR and its differentiable counterpart, we can analyze their differences in reverberation characteristics. Specifically, we obtain the EDR error $E^\text{EDR}$ by calculating the difference between the two log-magnitude EDRs.
            \begin{equation}
                E^\text{EDR} = 10\log_{10} \hat{H}^\text{EDR} - 10\log_{10} \tilde{H}^\text{EDR}.
            \end{equation}
            Figure \ref{fig:approx-error-fvn} shows the EDRs and the EDR error. It reveals that the error is small (at most $3\si{dB}$) and mostly in the low-energy region (below $-40\si{dB}$). Also, there is little EDR error at the first time index, i.e., $|E^\text{EDR}[k, 0]| \ll 1$, indicating that the overall magnitude response remains the same during the conversion process. In conclusion, the error is hardly noticeable, and the differentiable FVN can replace the original FVN for the training (we use a loss function related to the EDR; see Section \ref{subsection:loss}).
    \begin{figure}[t]
    \centering
    \subfloat[Modified FVN in the time domain. \label{fig:fvn-model}]{
        \includegraphics[width=.7188\columnwidth]{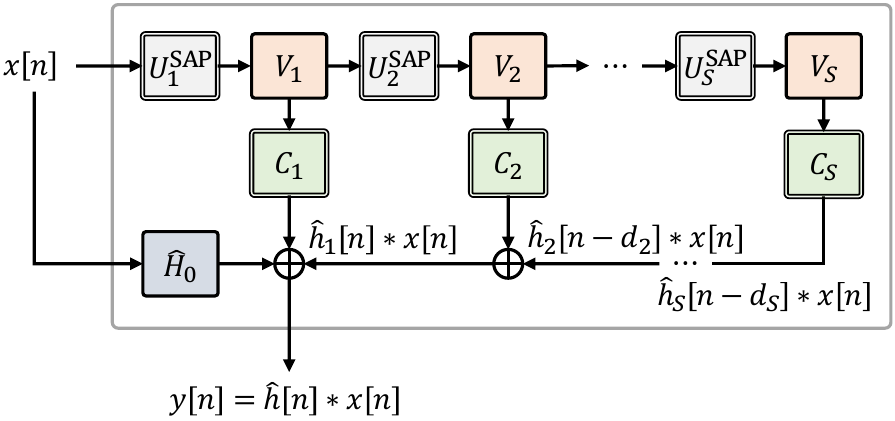}} \\
    \subfloat[Modified \begin{sc}AVFN\end{sc} in the time domain. \label{fig:advfvn-model}]{
        \includegraphics[width=.98\columnwidth]{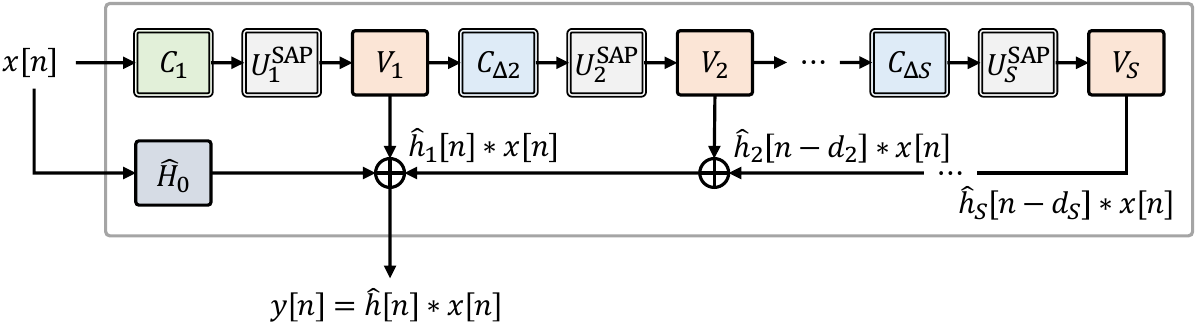}} \\
    \subfloat[Each velvet filter $V_i$. \label{fig:velvet-filter}]{
        \includegraphics[width=0.58\columnwidth]{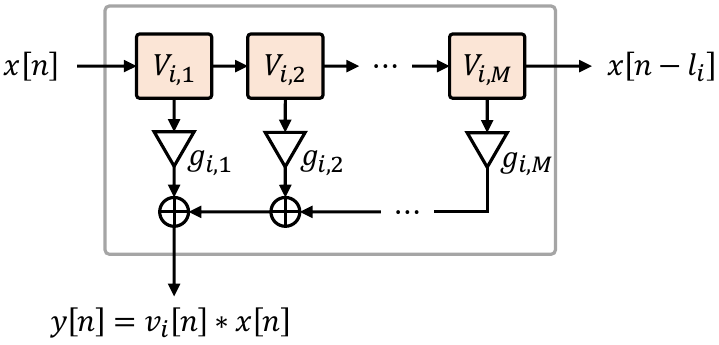}}
    \hfill
    \subfloat[Each coloration filter $C_i$. \label{fig:color-filter}]{
        \includegraphics[width=0.38\columnwidth]{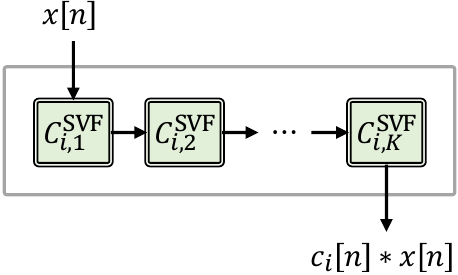}}
    \caption{
        Modifications to the original FVN and AFVN.
        (a) Modified FVN. To retain a sharp transient, we insert each SAP before its corresponding velvet filter. For each velvet filter $V_i$, the output denoted with a down arrow is a convolution of the input and the velvet noise $v_i[n]$. The other output denoted with a right arrow is a $L_i$-sample-delayed input. Refer to \cite{valimaki2017fvn} for more details. 
        (b) The same modification is performed to the \begin{sc}AVFN\end{sc} model. 
        (c) For both FVN and \begin{sc}AVFN\end{sc}, each velvet filter $V_i$ is divided into smaller filters $V_{i,1}, \cdots, V_{i, M}$ assigned with gains $g_{i, 1}, \cdots, g_{i, M}$. This modification is equivalent to dividing each velvet segment $v_i[n]$ into smaller segments $v_{i, 1}[n], \cdots, v_{i, M}[n]$ and multiplying them with $g_{i, 1}, \cdots, g_{i, M}$. 
        (d) The $i^\text{th}$ coloration filter $C_i$ in the time domain, which is serially connected SVFs $C_{i, 1}^\text{SVF}, \cdots, C_{i, K}^\text{SVF}$.} 
  \label{fig:fvn-2} 
\end{figure}
    \subsubsection{Modifications to the Original Model} \label{subsubsection:fvn-modifications}
    FVN was initially designed for the analysis-synthesis of late reverberation; it used the direct arrival and early reflections cropped from the RIR and modeled only the late part. However, we also aim to solve the blind estimation problem where we must estimate the entire RIR. To this end, we modify the original model as follows. 
    \begin{itemize}
        \item \emph{Finer segment gain.} We divide each velvet segment $v_i[n]$ into $M$ smaller segments $v_{i, 1}[n], \cdots, v_{i, M}[n]$ multiplied with gains $g_{i, 1}, \cdots, g_{i, M}$. Figure \ref{fig:velvet-filter} shows the resulting velvet filter $V_i$ in the time domain.
        \item \emph{Cumulative allpass.} The original model smoothes every velvet noise with the same allpass filter. This is undesirable in our context since it could smear the direct arrival and early reflections. Instead, we gradually cascade the SAPs to obtain each allpass filter $U_i(z) = \prod_{j=1}^i U_j^\text{SAP}(z)$ so the earlier segments are less smeared than the later ones. 
        As shown in Figure \ref{fig:fvn-model}, this modification can be implemented in the time domain by inserting each SAP $U_j^\text{SAP}$ before the corresponding velvet filter $V_j$.
        \item \emph{Shorter deterministic FIR.} With the above changes, the stochastic segments can fit the early RIR to some extent. Therefore, we shorten the early deterministic FIR $\hat{h}_0[n]$.
    \end{itemize}
    Refer to Appendix \ref{subsection:ablations} for the evaluation of the modifications.

    \subsubsection{Model Configuration Details} \label{subsubsection:FVN-configs}
        Exact configurations of our FVN used for the experiments are as follows. 
         
        We generated $2.5$ seconds of IR ($48\si{kHz}$ sampling rate, total $L=120\si{k}$ samples) with FVN. We used $S=20$ non-uniform velvet segments whose lengths $L_i$ are $ L/40, L/20$, and $L/10$ for $10, 5$, and $5$ segments, respectively, and their average pulse distances were set to $\mathbf{T} = [10$, $20$, $35$, $50$, $65$, $90$, $120$, $135$, $180$, $220$, $270$, $320$, $370$, $420$, $480$, $540$, $610$, $680$, $750$, $820]$. The SAPs $U_j^\text{SAP}$ were forced to have gains $\gamma_j = 0.75 + 0.01j$ and delay lengths $\boldsymbol{\tau}=[23$, $48$, $79$, $109$, $113$, $127$, $163$, $191$, $229$, $251$, $293$, $337$, $397$, $421$, $449$, $509$, $541$, $601$, $641$, $691]$. Each velvet segment had $M=4$ sub-segment gains and filtered with $K=8$ SVFs. We frequency-sampled the coloration filters with $N=4\si{k}$ points. The length of the deterministic FIR $h_0[n]$ was set to $50$. We set the gain $\mathbf{g}\in \mathbb{R}^{ S\times M}$, SVF parameters $\mathbf{f}, \mathbf{R}, \mathbf{m^\text{LP}}, \mathbf{m^\text{BP}},$ $\mathbf{m^\text{HP}}\in \mathbb{R}^{ S\times K},$ and the bypass FIR $\mathbf{h}_0 \in \mathbb{R}^Z$ to estimation {targets} (total $930$ ARPs). The time-domain FVN $\hat{H}$ requires $2166$ floating point operations per sample (FLOPs).

    \subsection{Differentiable Advanced Filtered Velvet Noise}
        Since frequency-dependent decay of reverberation is gradual, one can model each coloration filter $C_i$ as an initial coloration filter $C_1$ cascaded with delta-coloration filters $C_{\Delta 2}, \cdots, C_{\Delta i}$.
        \begin{equation}
            C_i(z) = C_1(z)\prod_{j=2}^{i} C_{\Delta j}(z).
        \end{equation}
        FVN with this modification is called \emph{Advanced Filtered Velvet Noise} ({AFVN}) \cite{valimaki2017fvn}.
        The delta filters' orders $\Delta K_2, \cdots, \Delta K_S$ are set lower than the initial filter's $K_1$ for the efficiency. See Figure \ref{fig:advfvn-model} for its time-domain implementation. 
        We used $K_1 = 8$ and $K_\Delta = K_{\Delta 2} = \cdots = K_{\Delta S} = 2$ in the experiments. This results in a total of $\Bar{K} = K_1 + (S-1)K_\Delta = 46$ SVFs. The other settings are the same as the FVN. This {AFVN} has $360$ ARPs to estimate and its time-domain model $\hat{H}$ requires $1511$ FLOPs. While order of the coloration filters are higher than the FVN's counterparts, the introduced frequency-sampling error remains approximately the same in practice (see Figure \ref{fig:approx-error-afvn}).

    \subsection{Differentiable Delay Network}
        \subsubsection{Delay Network} 
            By interconnecting multiple delay lines in a recursive manner, one can obtain a \emph{Delay Network} (DN) \cite{puckette1989reverb, gerzon1971reverb} structure.
            A difference equation of the DN can be written in a general form as follows (we omit the bracket notation for the filtering),
            \begin{subequations}
                \begin{align}
                    y[n] &= \mathbf{C}^T\Bar{\mathbf{y}}[n] + \hat{H}_0x[n], \\
                    \Bar{\mathbf{y}}[n + \mathbf{d}] &= \mathbf{A}\Bar{\mathbf{y}}[n] + \mathbf{B}x[n].
                \end{align}
            \end{subequations}
            That is, an input $x[n]$ is distributed and filtered (or simply scaled) with $\mathbf{B}$, then go through $\mathbf{d}$-sample parallel delay lines which are recursively interconnected to themselves through a mixing filter matrix $\mathbf{A}$. The delay line outputs $\Bar{\mathbf{y}}[n]$ are filtered and summed with $\mathbf{C}$. We add a bypass signal filtered with $\hat{H}_0$, resulting in an output $y[n]$. Transfer function of the DN is
            \begin{equation}
                \hat{H}(z) = \mathbf{C}(z)^T(\mathbf{D}^{-1}(z) - \mathbf{A}(z))^{-1}\mathbf{B}(z) + \hat{H}_0(z).
                \label{eq:tf-general-fdn}
            \end{equation}
            Here, $\mathbf{D}(z) = \text{diag} (z^{-\mathbf{d}})$ is an $M\times M$ transfer function matrix  for the delay lines, i.e., $\mathbf{D}_{ii}(z) = z^{-d_i}$. $\mathbf{B}(z)$, $\mathbf{C}(z)$, and $\mathbf{A}(z)$ are input, output, and feedback transfer function matrices of shape $M\times 1$, $1\times M$, and $M\times M$, respectively.
            
        \subsubsection{Differentiable Implementation}
            The DN components are recursively interconnected such that one cannot divide its IR into independent segments and generate them in parallel like we did with the FVN and {AFVN}. Instead, we frequency-sample its entire transfer function to obtain differentiable DN, which is equivalent to a composition of individually frequency-sampled transfer function matrices.
            \begin{equation}
                \Tilde{h}[n] = \text{IFFT}\left\{\mathbf{C}_N^T(\mathbf{D}_N^{-1} - \mathbf{A}_N)^{-1}\mathbf{B}_N + (\hat{H}_0)_N \right\}.
                \label{eq:diff-general-fdn}
            \end{equation}
            Here, $\mathbf{D}_N = \text{diag}((z^{-\mathbf{d}})_N) \in \mathbb{C}^{M\times M\times \left\lfloor N / 2 + 1 \right\rfloor}$ is a frequency-sampled version of the delay line transfer function matrix $\mathbf{D}$, i.e., $(\mathbf{D}_N)_{ii} = (z^{-d_i})_N \in \mathbb{C}^{\left\lfloor N / 2 + 1 \right\rfloor}$. Likewise, we can frequency-sample the transfer function matrices $\mathbf{B}$, $\mathbf{C}$, and $\mathbf{A}$ to obtain their approximations $\mathbf{B}_N$, $\mathbf{C}_N$, and $\mathbf{A}_N$, respectively. 
            Each frequency-sampled transfer function matrix is a batch of $\left\lfloor N/2 + 1 \right\rfloor$ matrices, the matrix multiplications and inversions of equation \eqref{eq:diff-general-fdn} can be performed in parallel.
        \begin{figure}[t]
    \begin{center}
        \centerline{\includegraphics[width=1\columnwidth]{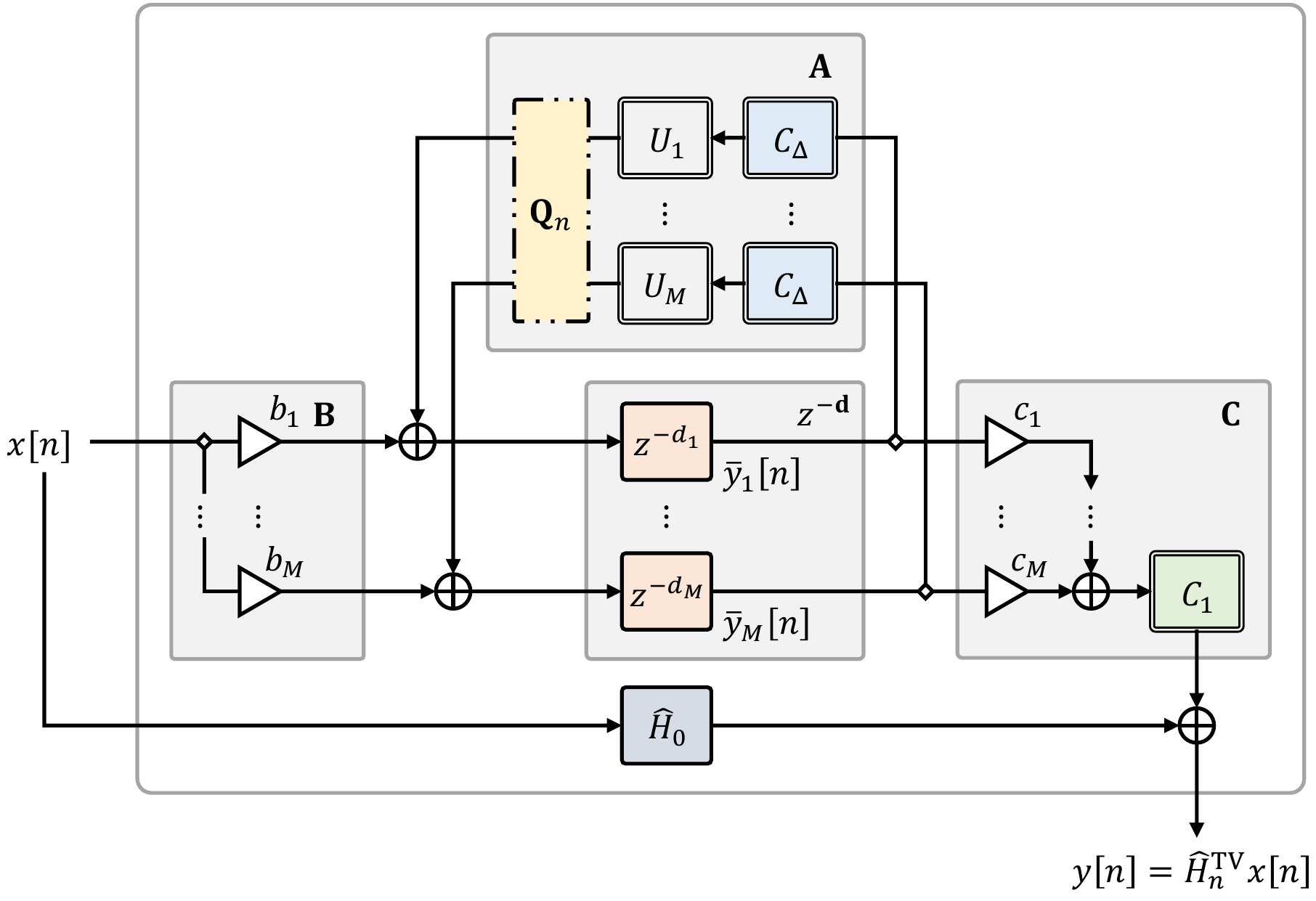}}
        \caption{Our DN in the time domain. 
        We restrict the general filter matrices to accelerate the training and obtain an efficient and controllable model. 
        We simplify the pre-filters $\mathbf{B}$ and the post-filters $\mathbf{C}$ to have only one coloration filter $C_1$ and use the same $C_\Delta$ for all feedback loops. A time-varying mixing matrix $\mathbf{Q}_n$ and allpass filters $\mathbf{U}$ are introduced to use DN with small $M$ (i.e., fast training) with minimal quality loss.}
        \label{fig:fdn-model}
    \end{center}
\end{figure}

        \subsubsection{Restrictions to the General Model} \label{subsubsection:fdn-modifications}
        The presented DN structure is fully general and able to express many variants, including the \emph{Feedback Delay Network} (FDN) \cite{jot1991fdn}. Therefore, we can also derive their differentiable versions following the proposed method.
        However, when following the general practice that uses many delay lines (e.g., $M=16$) for a high-quality reverberation, its differentiable model's frequency-sampled transfer function matrices consume too much memory, and their multiplications and inversions bottleneck the training speed. To tackle this, we reduce $M$ and modify the components as follows to make DN plausible with low $M$ (see Figure \ref{fig:fdn-model}).
        \begin{itemize}
            \item \emph{Pre and Post Filter Matrix.} Following most previous works \cite{rocchesso1997ellipticfdn, schlect2015time-varying, jot92fdnanalysissynthesis}, the pre filter matrix $\mathbf{B}$ is set to a (constant) gain vector $\mathbf{b}$. The post filter matrix $\mathbf{C}$ is a combination of a gain vector $\mathbf{c}$ and a common filter $C_1$, i.e., $\mathbf{C}=C_1\mathbf{c}$. The common filter $C_1$ is composed of serial $K_{C_1}$ SVFs.
            \item \emph{Feedback Filter Matrix.} While most DN model combines channel-wise parallel delta-coloration (absorption) filters $\mathbf{C}_\Delta$ and an inter-channel mixing matrix $\mathbf{Q}$ to compose the feedback filter matrix $\mathbf{A} = \mathbf{Q}\mathbf{C}_\Delta$. Additionally, we insert channel-wise allpass filters $\mathbf{U}$ and introduce \emph{time-variance} by modulating the mixing matrix $\mathbf{Q}$. This results in the time-varying feedback filter matrix $\mathbf{A}_n = \mathbf{Q}_n \mathbf{U}\mathbf{C}_\Delta$.
            
            \item \emph{Allpass Filters.} To achieve faster echo density build-up without adding more delay lines, we insert serial $K_\mathbf{U}$ SAPs in each feedback path.
            Unlike the FVN and {AFVN} cases, we estimate the SAP gains $\boldsymbol{\gamma} \in \mathbb{R}^{M\times K_\mathbf{U}}$ with the estimation network and frequency-sample the allpass filter matrix $\mathbf{U}$ for the differentiable model.
            \item \emph{Time-varying Mixing Matrix.} 
            We further reduce the audible “ringing" artifacts by modulating the stationary poles with the time-varying mixing matrix, which is set to $\mathbf{Q}_n = \mathbf{Q}_0 \mathbf{R}^n$ where $\mathbf{Q}_0$ is a  Householder matrix and $\mathbf{R}$ is a tiny rotational matrix constructed from a random matrix \cite{schlect2015time-varying, sebastian2020fbntb}. Since the frequency-sampling method is only applicable to LTI filters, we fix the mixing matrix to $\mathbf{Q}_0$ when obtaining the differentiable model. 
            \item \emph{Absorption Filters.} 
            Unlike the conventional FDN \cite{jot1991fdn}, we use the same absorption filter $C_\Delta$ for every channel.
            Since the mixing matrix $\mathbf{Q}_n$ is always unitary, our DN becomes stable when $C_\Delta$ has magnitude response less than $1$ \cite{schlect2015time-varying}. 
            We achieve this by using a PEQ with additional constraints (see Section \ref{subsection:activ-init}) as $C_\Delta$. 
            \item \emph{Bypass FIR.} We use a length-$Z$ FIR $h_0[n]$ for the bypass. 
        \end{itemize}
        
        \begin{figure}[t]
    \begin{center}
        \centerline{\includegraphics[width=\columnwidth]{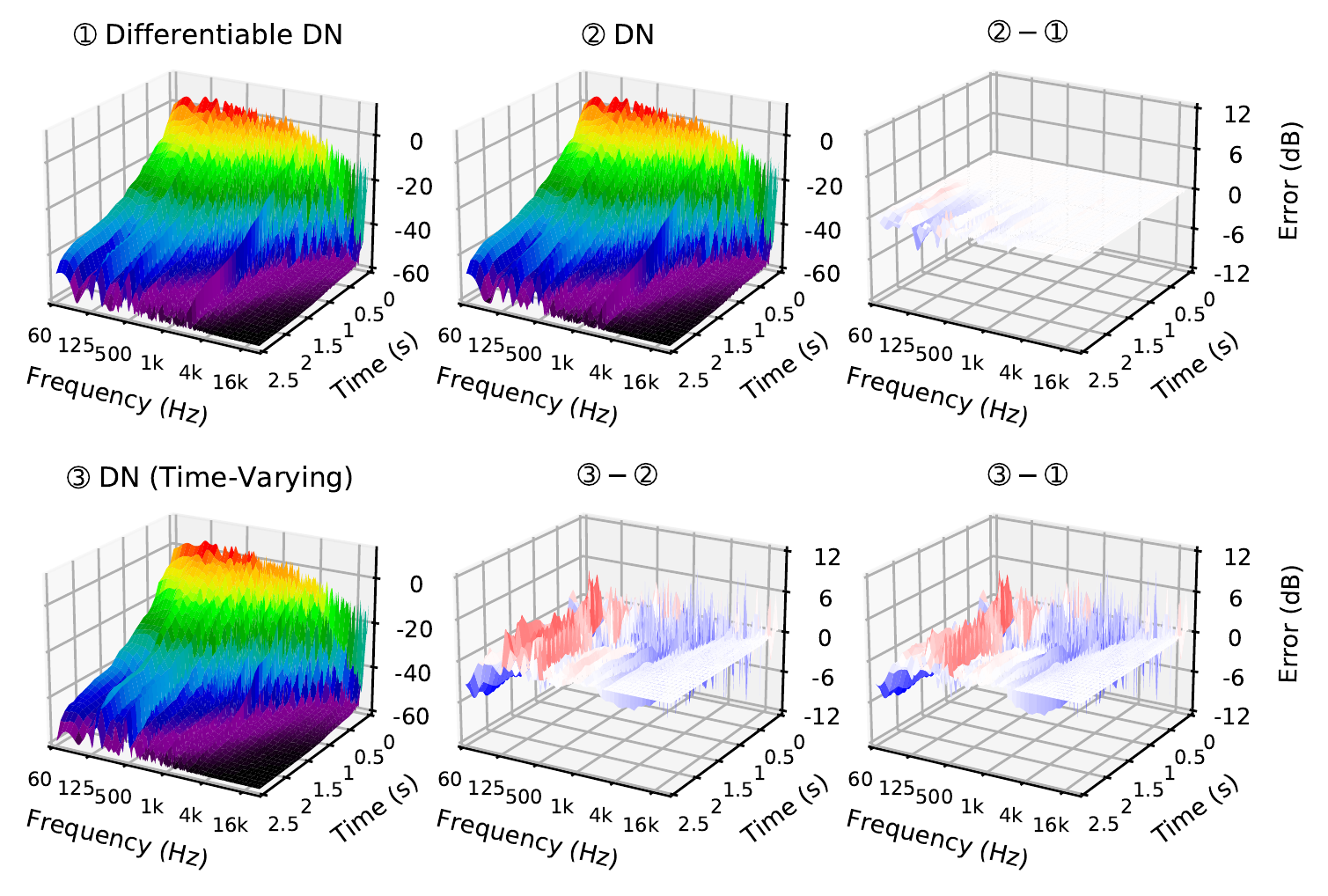}}
        \caption{EDR of differentiable DN, its original DN with two versions (LTI and time-varying), and EDR errors between those three. Same as the FVN and AFVN, the frequency-sampling introduce little error, as shown in the upper right plot. However, rotating the mixing matrix introduces a considerable amount of EDR error since it modulates the pole positions. The error is mostly in the low-frequency region, where the poles are sparsely located. 
        } 
        \label{fig:approx-error-fdn}
    \end{center}
\end{figure}

        We summarize the modifications. The difference equation of our time-varying DN $\hat{H}_n^\text{TV}$ is
        \begin{subequations}
            \begin{align}
                y[n] &= C_1\mathbf{c}^T\Bar{\mathbf{y}}[n] + H_0x[n], \\
                \Bar{\mathbf{y}}[n+\mathbf{d}] &= \mathbf{Q}_n\mathbf{U}\mathbf{C}_\Delta\Bar{\mathbf{y}}[n] + \mathbf{b}x[n].
            \end{align}
        \end{subequations}
        The transfer function of its LTI approximation $\hat{H}(z)$ is given as equation \eqref{eq:tf-general-fdn} with the restrictions $\mathbf{B}(z) = \mathbf{b}$, $\mathbf{C}(z) = C_1(z)\mathbf{c}$, and $\mathbf{A}(z) = \mathbf{Q}_0\,\text{diag}(\mathbf{U}_N(z)\odot\mathbf{C}_\Delta(z))$.
        Hence, the IFFT of 
        \begin{equation}
           (C_1)_N \mathbf{c}^T(\mathbf{D}_N^{-1} - \mathbf{Q}_0\,\text{diag}(\mathbf{U}_N\odot(\mathbf{C}_\Delta)_N))^{-1}\mathbf{b}
        \end{equation}
        summed with the bypass FIR $\hat{h}_0[n]$ results in our differentiable DN's IR $\Tilde{h}[n]$.
        
        Figure \ref{fig:approx-error-fdn} shows the EDR errors between the three different DN modes, i.e., differentiable DN, original LTI, and the time-varying one. The only difference between the first two is the frequency-sampling; unless the DN decays very slowly, there is little time-aliasing and EDR error. However, due to the pole modulation, the EDR error increases significantly when the time-varying mixing matrix is applied. Despite this, we find that other perceptually important factors, such as reverberation time, remain largely unchanged (see Section VI-A).
        \subsubsection{Model Configuration Details} We used $M=6$ delay lines with delay lengths $\mathbf{d}=[233$, $311$, $421$, $461$, $587$, $613]$ in samples. 
        We used $K_\mathbf{U} = 4$ SAPs for each delay line. We fixed the SAP delay lengths to $\boldsymbol{\tau}=[[131,$ $151,$ $337,$ $353],$ $[103,$ $173,$ $331,$ $373],$ $[89,$ $181,$ $307,$ $401],$ $[79,$ $197,$ $281,$ $419],$ $[61,$ $211,$ $257,$ $431],$ $[47,$ $229,$ $251,$ $443]]$. {Note that the SAPs introduce additional delays so that the effective delay lengths are $[1205$, $1291$, $1399$, $1437$, $1547$, $1583]$ or about $[25.1$, $26.9$, $29.1$, $29.9$, $32.2$, $33.0]$ in milliseconds.}
        Our rotational matrix satisfies $\mathbf{R}^{30\si{k}} = \mathbf{I}$ so that the mixing matrix $\mathbf{Q}_n$ has a period of $0.625$ second. Both post $C_1$ and absorption filter $C_\Delta$ have $K_{C_1}=K_{c_\Delta}=8$ components. We used $N = 120\si{k}$ frequency-sampling points. 
        The order of the bypass FIR is $Z=100$. The post filter parameters $\mathbf{f}_{C_1}$,  $\mathbf{R}_{C_1}$, $\mathbf{m}^\text{LP}_{C_1}$, $\mathbf{m}^\text{BP}_{C_1}$,  $\mathbf{m}^\text{HP}_{C_1} \in \mathbb{R}^{1\times K_{C_1}}$, feedback filter parameters $\mathbf{f}_{C_\Delta}$, $ \mathbf{R}_{C_\Delta}$, $\mathbf{G}_{C_\Delta} \in \mathbb{R}^{K_{C_\Delta}}$, pre and post gain vectors $\mathbf{b}$, $\mathbf{c} \in \mathbb{R}^{M\times 1}$, SAP gain vector $\boldsymbol{\gamma} \in \mathbb{R}^{M \times K_\mathbf{U}}$, and the bypass FIR $\mathbf{h}_0\in \mathbb{R}^{Z}$ are the estimation {targets} (total $200$ ARPs). The resulting time-domain model $\hat{H}$ and the time-varying model $\hat{H}^\text{TV}_n$ consume $889$ and $1285$ FLOPs, respectively.

\section{Artificial Reverberation Parameter Estimation with a Deep Neural Network}\label{section:arp-estimation}
    The details of our ARP estimation network are as follows. As shown in Figure \ref{fig:network-archi}, it transforms single channel audio input (either RIR or reverberant speech) into a shared latent $\mathbf{z}$ with the model/task-agnostic encoder. Then, each ARP-groupwise layer projects the latent $\mathbf{z}$ into an ARP tensor $\mathbf{P}_i$. 
    
    \subsection{AR-Model/Task-Agnostic Encoder}
\begin{figure}[t]
    \begin{center}
        \centerline{\includegraphics[width=.86\columnwidth]{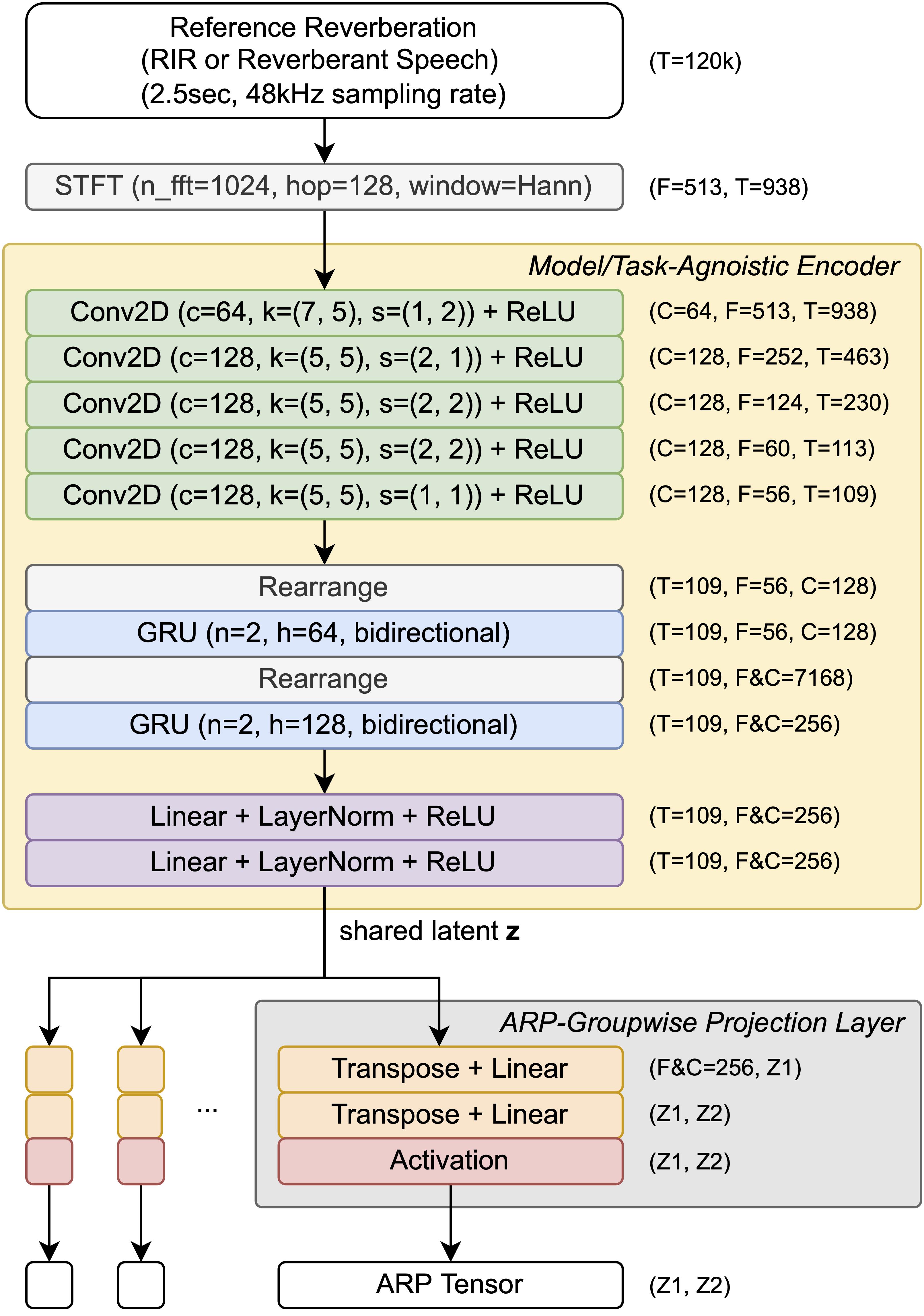}}
        \caption{Architecture of the ARP estimation network. For each two-dimensional convolutional (Conv2D) layer, $c$, $k$, and $s$ denote number of output channel, kernel size, and stride, respectively. For each gated reccurent unit (GRU) layer, $n$ and $h$ denote number of layers and hidden features. Shape and size of the intermediate output tensors are provided in the parentheses, where $F$, $T$, and $C$ denote frequency, time, and channel axis, respectively. $Z1$, and $Z2$ denote each ARP tensor's first and second axis. We omitted the batch axis.}
        \label{fig:network-archi}
    \end{center}
\end{figure}

        We first transform the reference RIR or reverberant speech into a log-frequency log-magnitude spectrogram. Then, the spectrogram goes through five two-dimensional convolutional layers, each followed by a rectified linear unit (ReLU) activation. We apply gated recurrent unit (GRU) layers \cite{chung2014empirical} along with the frequency axis regarding the channels as features and the time axis regarding the channels and frequencies as features, widening the receptive field. Finally, we apply two linear layers followed by layer normalization \cite{ba2016layer} and ReLU, resulting in the two-dimensional shared latent $\mathbf{z}$. The detailed network configurations are shown in Figure \ref{fig:network-archi}, and the encoder has about $7.3\si{M}$ parameters. 

    \subsection{ARP-Groupwise Projection Layers}
        Now we aim to transform the two-dimensional shared latent $\mathbf{z}$ into each ARP tensor $\mathbf{P}_i$ which is at most two-dimensional. 
        To achieve this, we use two axis-by-axis linear layers. We first transpose and apply a linear layer to the latent to match the first axis shape. Then, another transposition and a linear layer match the other axis. This approach factorizes a single large layer into two smaller ones, reducing the number of parameters.

    \subsection{Nonlinear Activation Functions and Bias Initialization}
        \label{subsection:activ-init}
        Each ARP group has different stability conditions and desired distribution. To satisfy these, we attach a nonlinear activation to each projection layer and initialize the last linear layer's bias as follows. Here, $x$ denotes a pre-activation element.
        
        \subsubsection{Nonlinear Activation Functions}
        \begin{itemize}
            \item For the resonance $\mathbf{R}$ of the SVF, we use a scaled softplus $\zeta(x) = \log(1 + e^x)/\log(2)$ to center the initial distribution at $1$ and satisfy the stability condition $\mathbf{R}>\mathbf{0}$. 
            \item For cutoff $\mathbf{f}$ {we} also have the same stability condition $\mathbf{f}>\mathbf{0}$. However, instead of the softplus, we use $\tan(\pi \sigma(x)/2)$ where $\sigma(x) = 1/(1+e^{-x})$ is a logistic sigmoid. With this activation, $\sigma(x)=0$ and $1$ represent cutoff frequency of $0\si{Hz}$ and half of the sampling rate, respectively. 
            \item The delta-coloration PEQ $C_\Delta$ inside the DN must satisfy the stability condition $|C_\Delta(e^{j\omega})| < 1$. To achieve this, we use $10^{-\zeta(x)}$ for the component gain $\mathbf{G}$.
            Also, the shelving filters' resonances should satisfy $\mathbf{R}>\sqrt{2}/2$ to prevent their magnitude responses to ``spike'' over $1$. Therefore, we add $\sqrt{2}/2$ after the softplus activation.
            \item For the feed-forward/back gains $\boldsymbol{\gamma}$ of channel-wise allpass filters $\mathbf{U}$ inside the DN, we use sigmoid activation $\sigma(x)$.
        \end{itemize}
        
        \subsubsection{Bias Initialization}
        \begin{itemize}
            \item We initialize bias of the $\mathbf{f}$ projection layer to $\sigma^{-1}(2\omega_k/\omega_s)$ where each $\omega_k$ is desired initial cutoff frequency of $k^\text{th}$ SVF. We set $\omega_k = \omega_\text{min}(\omega_\text{max} / \omega_\text{min})^{(k-1)/(K-1)}$ such that the frequencies are equally spaced in the logarithmic scale from $\omega_\text{min}=40\si{Hz}$ to $\omega_\text{max}=12\si{kHz}$. To the SVFs inside the {AFVN}'s initial and delta-coloration filters, we perform a random permutation to the frequency index $k$ and obtain a new index $\Bar{k}$. This prevents SVFs for the latter segments having higher initial cutoff frequencies.
            \item Decoder biases for the SVF mixing coefficients $\mathbf{m}^\text{LP}, \mathbf{m}^\text{BP}$, and $\mathbf{m}^\text{HP}$ are initialized to $1, 2$, and $1$, respectively, so that each SVF's initial magnitude response slightly {deviates} from $1$. This prevents the coloration filters' responses and the estimation network loss gradients to vanish or explode.
            \item For the PEQ gains $\mathbf{G}$ of the DN's delta-coloration filters, we set each of their biases to $-10$ such that the initially generated IRs have long enough reverberation time.    
        \end{itemize}
        
    \subsection{Loss Function} \label{subsection:loss}
        \subsubsection{Match Loss}
            We utilize a multi-scale spectral loss \cite{engel2020ddsp} and define a match loss as follows,
            \begin{equation}
                \mathcal{L}_\text{\begin{sc}Match\end{sc}}(h, \Tilde{h}) = \sum_i \left\| |H^\text{STFT}_i| - |\tilde{H}^\text{STFT}_i| \right\|_1.
            \end{equation} $i$ denotes STFT with different FFT sizes. We use the FFT sizes of $256$, $512$, $1024$, $2048$, and $4096$, hop sizes of $25\%$ of the respective FFT sizes, and Hann windows. Each spectrogram's frequency axis is log-scaled like the encoder's spectrogram. 
            
        \subsubsection{Regularization}
            We additionally apply a regularization to reduce time-aliasing of the estimation.
            As shown in equation \eqref{eq:fr-deviation} and \eqref{eq:deviation-loss}, pole radii of each SVF affects the reliability of the frequency-sampled one (hence DAR model) and its parameter estimator. Since each SVF's IR $(c^\text{SVF}_{i, k})_N[n]$ can be computed, we encourage reducing its pole radii by penalizing the IR's decreasing speed. Each decreasing speed $\gamma_{i, k}$ is calculated by the ratio of the average amplitude of first and last $n_0$ samples of $(c^\text{SVF}_{i, k})_N[n]$. We fix $n_0$ to $N/8$. Then, each $\gamma_{i, k}$ is weighted with a softmax function along the SVF-axis to penalize higher $\gamma_{i, k}$ more. Sum of the weighted decreasing speed values results in the regularization loss $\mathcal{L}_\text{\begin{sc}Reg\end{sc}}$ as follows,\begin{subequations}
                \begin{align}
                    \gamma_{i, k} &= \frac{\sum_{n=N-n_0}^N | (c^\text{SVF}_{i, k})_N[n] | }{\sum_{n=0}^{n_0} | (c^\text{SVF}_{i, k})_N[n] |}, \\
                    \mathcal{L}_\text{\begin{sc}Reg\end{sc}} &=  \sum_{i=1}^S\frac{\sum_{k=1}^K \gamma_{i, k} e^{\gamma_{i, k}}}{\sum_{j=1}^K e^{\gamma_{i, j}}}.
                \end{align}
            \end{subequations}
            We omit the regularization loss for the DN networks since with DN the time-aliasing is inevitable to match the {reference} with reverberation time longer than the number of frequency-sampling points $N$. 
            Therefore, our full loss function $\mathcal{L}$ is
            \begin{equation}
                \mathcal{L}(h, \Tilde{h}) = \mathcal{L}_\text{\begin{sc}Match\end{sc}}(h, \Tilde{h}) + \beta \mathcal{L}_\text{\begin{sc}Reg\end{sc}}(\mathbf{c}^\text{SVF}_N)
                \label{eq:loss}
            \end{equation}
            where $\beta = 1$ for the FVN and {AFVN} and $\beta = 0$ for the DN.

\section{Experimental Setup} \label{section:experimental-setup}
    \subsection{Data} \label{subsection:data}
        \subsubsection{Room Impulse Response}
            We collected $1835$ real-world RIR measurements from various datasets including OpenAIR \cite{simon2010openair} and ACE Challenge \cite{eaton2016ace}.
            The amount of the RIRs were insufficient for the training purpose, so we used them only for the validation and test ($836$ and $999$ RIRs, respectively), ensuring that each set consists of RIRs from different rooms. 
            
            For the training, we synthesized $200\si{k}$ RIRs with shoebox room simulations using the image-source method \cite{svensson2002computational, pyroom}. To reflect various acoustic environments, we randomized simulation parameters for every RIR synthesis, which include room size, each wall's frequency-dependent absorption coefficient, and source/microphone positions. We tuned the randomization scheme to match the training dataset to the validation set in terms of reverberation parameter statistics.

            In addition, we pre-processed each RIR as following. 
            \begin{itemize}
                \item We removed the pre-onset part of the RIR. 
                We detected the onset by extracting a local energy envelope of the RIR then finding its maximum point \cite{guillaume08onset}.
                \item Following the recent RIR augmentation method \cite{9052970}, we multiplied random gain sampled from $\mathcal{U}(-12\si{dB}, 3\si{dB})$ to the first $5\si{ms}$ of each train RIR. This slightly reduces the average direct-to-reverberant ratio ($\text{DRR}$) of the training set and matches the validation set.
                At the evaluation/test, we omitted this procedure.
                \item Finally, we normalized the RIR to have the energy of $1$. 
            \end{itemize}

        \subsubsection{Reverberant Speech}
            We used VCTK \cite{vctk} for dry speech. We split it into the train ($61808$), validation ($21608$), and test ($4912$) set so that each set is composed of the speech recordings from different speakers. 
            We sampled an RIR and a dry speech sample from their respective datasets and convolved them to generate reverberant speech. We random-cropped $2.5$-second segment from it for the input.

    \subsection{Training}
        We trained each network with every proposed DAR model for the three tasks: analysis-synthesis, blind estimation, and both tasks. For the both-performing ones, we fed an RIR or reverberant speech in a $50$-$50\%$ probability.
    
        We set the initial learning rate to $10^{-4}$ for FVN and {AFVN} and $10^{-5}$ for DN. We used Adam optimizer \cite{kingma2017adam}. For DN, gradients were clipped to $\pm10$ for stable training. After $250\si{k}$ steps, we performed the learning rate decay. We multiplied {the learning rate by} $10^{-0.2}$ and $10^{-0.1}$ every $50\si{k}$ step for the analysis-synthesis networks and the others, respectively.
        We finished the training if the validation loss {did not improve after $50\si{k}$ steps}, which took no more than $500\si{k}$ steps for the analysis-synthesis and $1\si{M}$ steps for the others. 

    \subsection{Evaluation Metrics}
        We evaluated our networks with the match loss $\mathcal{L}_\text{\begin{sc}Match\end{sc}}$ and EDR distance, defined as an average of the absolute EDR error. Furthermore, we evaluated reverberation parameter differences, i.e., reverberation time $\text{T}_\text{30}$, $\text{DRR}$, and clarity $\text{C}_\text{50}$ difference, denoted as $\Delta \text{T}_\text{30}$, $\Delta \text{DRR}$, and $\Delta \text{C}_\text{50}$, respectively \cite{2008ISO3382_2, Hak2012decay}. We calculated both full-band differences and average of differences measured at octave bands of center frequencies of $125$, $250$, $500$, $1\si{k}$, $2\si{k}$, $4\si{k}$ and $8\si{kHz}$. Then, we obtained their respective median values.

        We can interpret the reverberation parameter differences by comparing them with their respective just noticeable differences \cite{agus2018minimally}. Reported just noticeable differences of $\text{T}_\text{30}$ from previous works vary from $5\%$ \cite{meng2006jnd_rt} up to about $25\%$ \cite{blevins13jndrt}. For $\text{C}_\text{50}$, $1\si{dB}$ \cite{BRADLEY1999c50jnd} and $1.1\si{dB}$ \cite{martellotta2020jnd_c_ts} were reported. For $\text{DRR}$ it differs for various range, e.g., $6\si{dB}$ at $-10\si{dB}$, $2\si{dB}$ at $0\si{dB}$,  $3\si{dB}$ at $10\si{dB}$, and $8\si{dB}$ at $20\si{dB}$ \cite{erik2008drrjnd}. 

    \subsection{Subjective Listening Test}
        Following the previous researches \cite{sarroff20revmatch, steinmetz2021filtered}, we conducted a modified MUltiple Stimuli with Hidden Reference and Anchor (MUSHRA) test \cite{mushra}.
        We asked subjects to score similarities of reverberation between a given reference reverberant speech $h * x$ and followings ($h$ and $x$ denote a test RIR and dry speech signal, respectively).
        \begin{itemize}
            \item A hidden reference $h * x$ (exactly the same audio).
            \item A lower anchor $h' * x$ obtained by averaging the test set RIRs \cite{steinmetz2021filtered}.
            \item Another anchor $h'' * x$ obtained by random-sampling an RIR from the test set \cite{sarroff20revmatch}.
            \item Estimations to evaluate $\hat{H}_1(x), \cdots, \hat{H}_n(x)$. We used the AR models $\hat{H}_1, \cdots, \hat{H}_n$ to obtain the reverberant speech. 
        \end{itemize}
        The similarity score ranged from $0$ to $100$, where we provided text descriptions for $5$ equally-divided ranges ($0$-$20$: totally different, $20$-$40$: considerably different, $40$-$60$: slightly different, $60$-$80$: noticeable, but not much, and $80$-$100$: imperceptible). 
        While we asked the participants to score both analysis-synthesis and blind estimation results, we divided them into two separate pages. As a result, a total of $10$ stimulus ($3$ proposed models, $4$ baselines which we will explain in Section \ref{subsection:baseline}, $2$ anchors, and $1$ hidden reference) were scored for each task.
        We conducted the test using the webMUSHRA API \cite{schoeffler2018webmushra}. A total of $24$ sets were scored for each task, where the first $3$ sets were used for training. A total of $12$ people participated where $7$ people were audio engineers and the others were instrument players. 
        We discarded $1$ subject's results since he scored the hidden references  lower than $90$ for more than $15\%$ of the entire sets.
    
    \begin{figure}[!t]
    \centering
    \subfloat[ARP estimation framework without the DAR models. \label{fig:baseline-param-match}]{
        \includegraphics[width=\columnwidth]{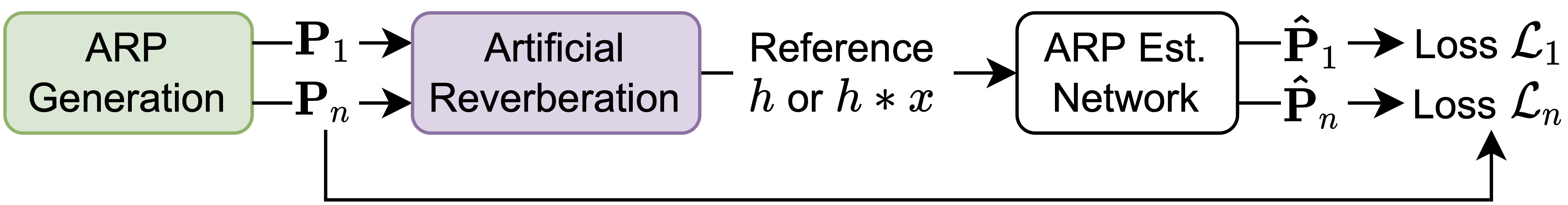}} \\
    \subfloat[RIR estimation with CNN decoder model. \label{fig:baseline-cnn}]{
        \includegraphics[width=0.77\columnwidth]{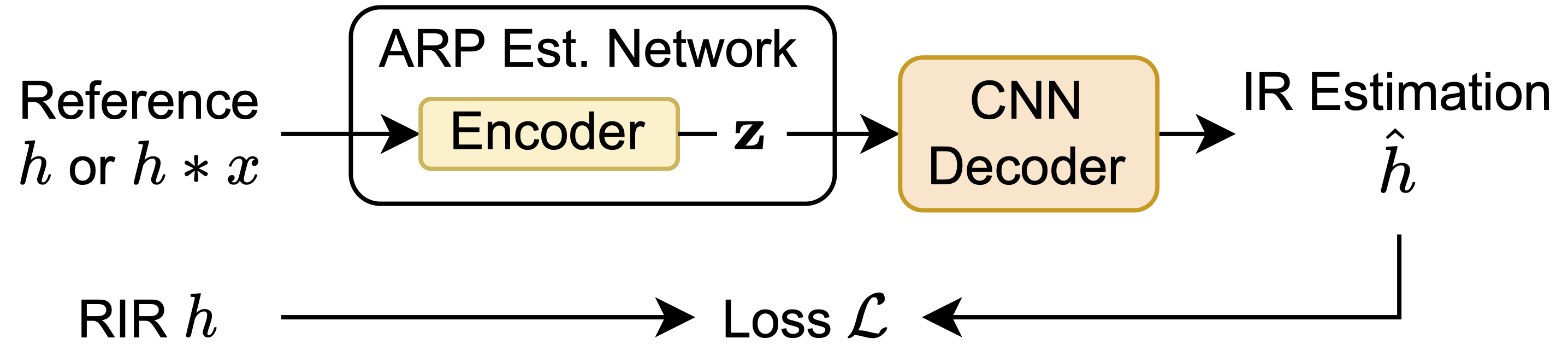}} \\
    \caption{Baseline methods for the experiments.
    }
  \label{fig:baseline-framework} 
\end{figure}
    \subsection{Baselines} \label{subsection:baseline}
        We compared our framework with following baselines. Refer to Appendix \ref{appendix:parameter-match-baseline} for further details.
        \subsubsection{Parameter-matching Networks}
            Assuming that the DAR models are not available, similar to the previous non-end-to-end approach \cite{sarroff20revmatch}, we generated the training data (reference-and-ARPs pair) with the DAR models and trained the same networks with a ARP-matching loss (see Figure \ref{fig:baseline-param-match}). Note that we used the DAR models simply for the convenience of on-the-fly data generation in GPU, but the same procedure can be performed with the AR models. 
            
        \subsubsection{CNN Decoder}
            We trained the same estimation network but instead of the DAR model we attached a decoder composed of one-dimensional transposed convolutional layers and TCNs. With this decoder, the entire network resembles an autoencoder (see Figure \ref{fig:baseline-cnn}). From now, we denote this model as CNN. {We trained two CNNs, one for each of the two tasks.}
            
        Note that we did not compare the our methods with possible two-stage solutions \cite{2008ISO3382_2, rama2003rtestimate, wen2008rtestimate, gamper2016rtestimatecnn, li2019freqrt_estimate, sebastian17revtimecontrol, karolina19imprevtimecontrol, chourdakis2017a} (see Section \ref{subsection:two-stage}) since each DAR model's architecture became different from its original.
\section{Evaluation Results}\label{section:evaluation-results}
    \subsection{Comparison with the Baseline Models}
        \subsubsection{Advantage of the End-to-end Learning}
\begin{figure}[t]
    \begin{center}
        \centerline{\includegraphics[width=\columnwidth]{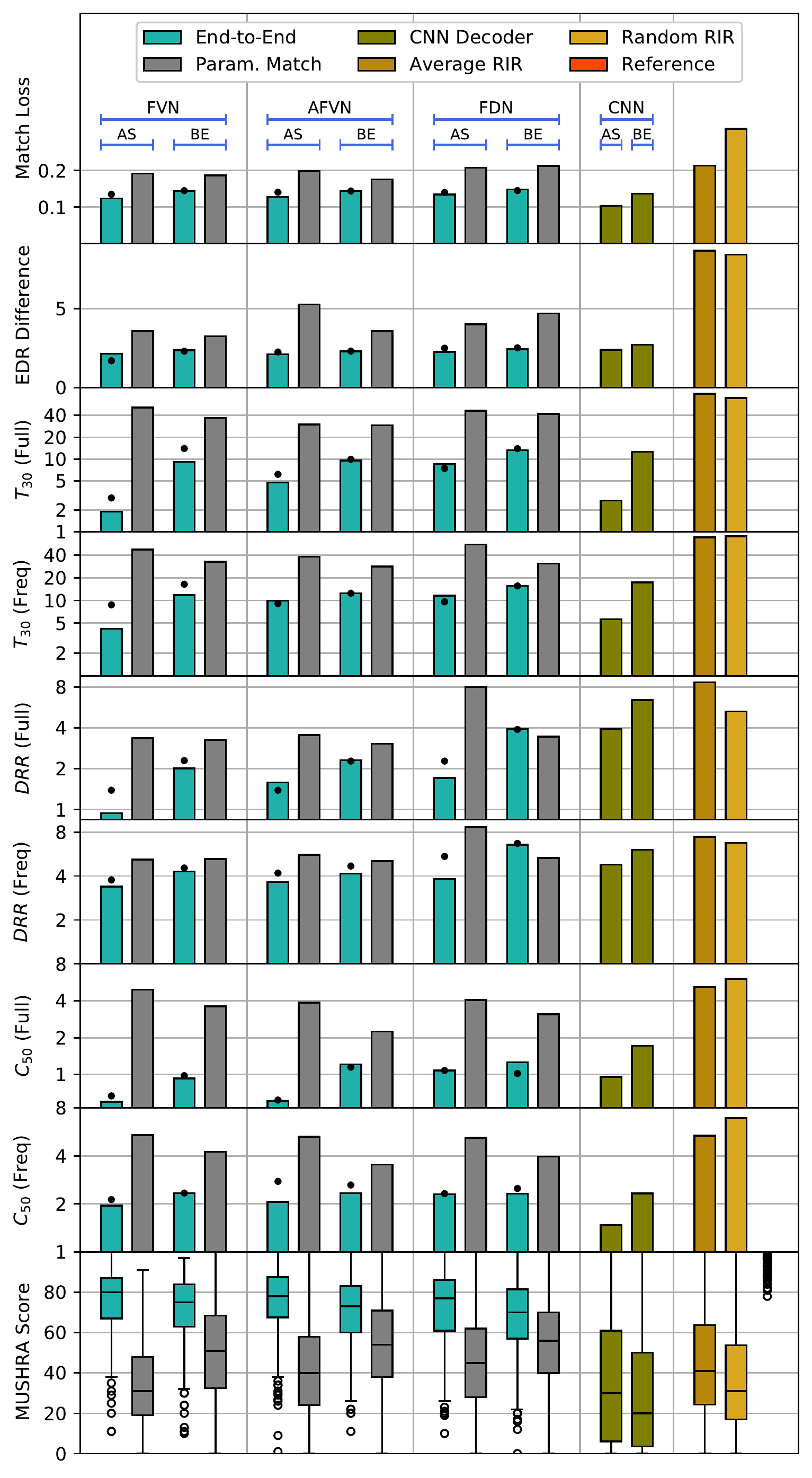}}
        \caption{ARP estimation results on various tasks with the AR/DAR models. Here, \texttt{AS} and \texttt{BE} denote analysis-synthesis and blind estimation, respectively. The \texttt{Full} and \texttt{Freq} denote the full-band reverberation parameter difference and average of octave-band reverberation parameter differences, respectively. Along with the proposed approaches and baselines, we additionally report the lower anchor's evaluation results and the (hidden) reference's MUSHRA score result. The both-performing networks' performance are denoted as small dots in the figure.}
        \label{fig:result}
    \end{center}
\end{figure}

        Figure \ref{fig:result} compares the proposed end-to-end approach with the baselines on objective metrics and subjective scores.
        By a large margin, the end-to-end model outperformed the non-end-to-end ARP-matching baselines. The baselines struggled to match the reverberation decay, reporting noticeably large reverberation time differences $\Delta T_{30} \geq 20\%$. 
        A two-sided t-test for each AR/DAR model and task showed that the performance difference was statistically significant (all $p<10^{-5}$).   
        Subjective listening test results also agreed with the objective metrics; median scores of the end-to-end approaches were from $70$ to $80$, showing close matches to the reference, while the parameter matching models scored less than $60$.
        Also, refer to Figure \ref{fig:edr} that visualizes each estimation result with an EDR and EDR error plots. 
        This shows that the end-to-end learning enabled by the DAR models gives a huge performance gain relative to the non-end-to-end parameter-matching scheme.

        \subsubsection{Advantage of the DSP Prior}
        Since the CNN baselines have more trainable parameters and complex architecture, they reach lower losses than the DAR-equipped networks. However, they produce audible artifacts. While being more restrictive, the AR models avoid such artifacts and achieve higher subjective listening scores than the CNN baselines. Again, these subjective score differences were statistically significant (all $p<10^{-5}$). 
        
    \begin{figure}[t]
    \begin{center}
        \centerline{\includegraphics[width=\columnwidth]{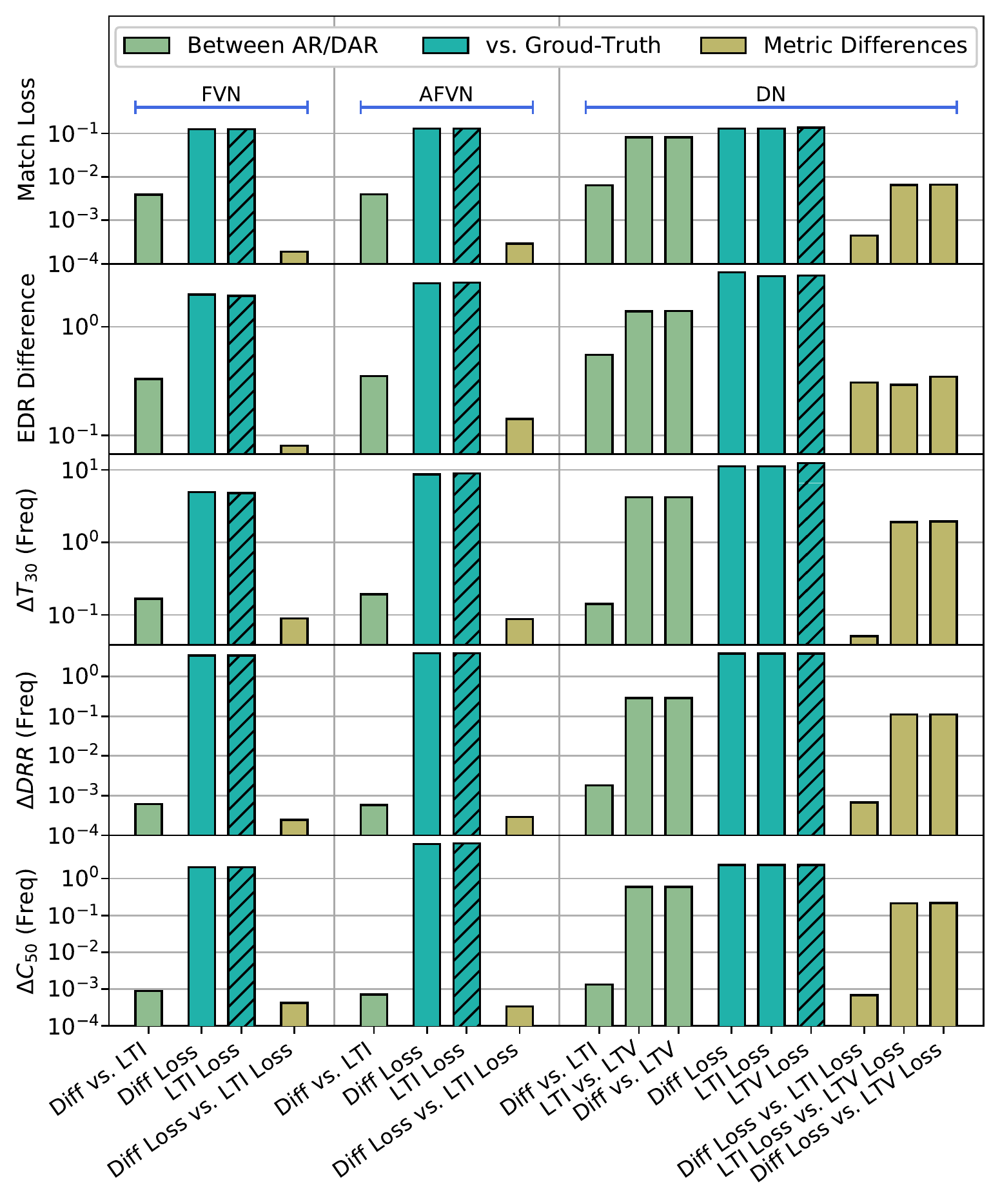}}
        \caption{Reliability of the DAR models. \texttt{Diff}, \texttt{LTI}, and \texttt{LTV} denote the differentiable, original LTI, and the time-varying model, respectively. The hatched results are the same values of Figure \ref{fig:result} (analysis-synthesis).}
        \label{fig:ar-dar-compare}
    \end{center}
\end{figure}

    \begin{figure}[t]
    \begin{center}
        \centerline{\includegraphics[width=0.98\columnwidth]{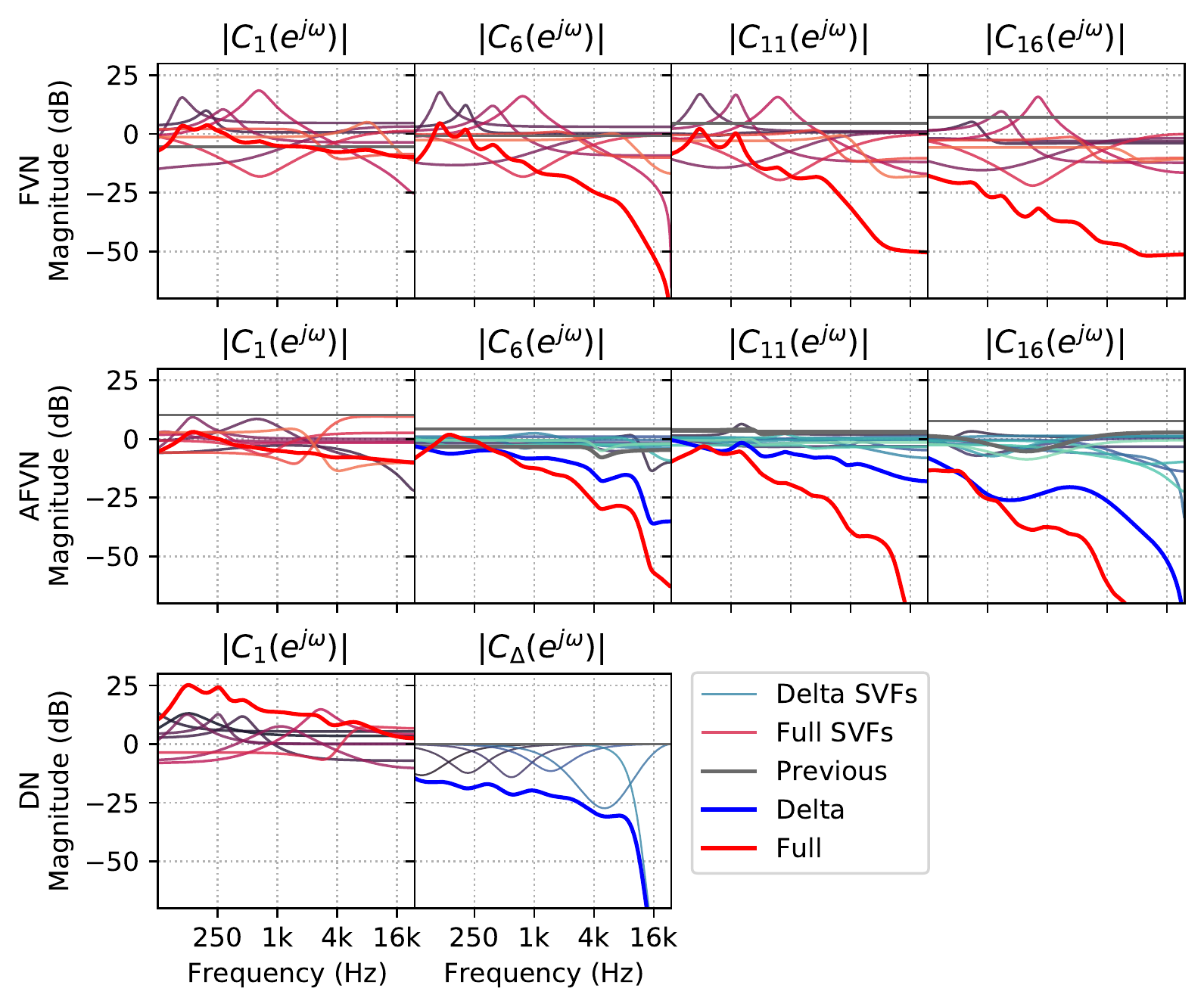}}
        \caption{
        Magnitude responses of the coloration filters of the AR models. 
        We used the proposed analysis-synthesis networks to obtain the filter parameters for this plot. 
        Each AR model has a different approach to model the frequency-dependent characteristic of given reverberation.
        }
        \label{fig:magnitude-response}
    \end{center}
\end{figure}    

    \subsection{Reliability of the DAR Models}
        We demonstrated in previous sections that each DAR model is close to the original AR model, making them reliable to use as an alternative for the training. Here, we empirically validate this argument again. Figure \ref{fig:ar-dar-compare} summarizes the objective metrics calculated with the DAR and AR models. We used the trained analysis-synthesis network for this evaluation. First, metrics calculated between the AR model IR and the DAR model IR were orders of magnitude smaller than the metrics calculated with the ground truth. DN was the only exception, showing more deviation than the other AR models due to the time variance. Nevertheless, the reverberation parameter differences between the DN IR and differentiable model IR were less than the just noticeable difference values (for example, $\Delta T_{30} < 3\%$). Second, as expected with equation \eqref{eq:loss-bound}, each metric difference was smaller than the corresponding AR-to-DAR conversion error. As a result, overall evaluation results were very similar regardless of the used model. In short, the DAR models are reliable for training in practice. 

    \subsection{Performance Difference Between {Target} Tasks}
        Without surprise, the analysis-synthesis networks performed better in most metrics than the blind estimation networks since the {reference} reverberation is directly given, i.e., the former ones have fewer burdens than the latter.
        Likewise, the both-performing networks (shown as dots in Figure \ref{fig:result}) show slightly degraded results than their single-task counterparts. Yet, the reverberation parameter difference only slightly increased (less than the just noticeable differences), which hint at a possibility of a universal reference-form-agnostic ARP estimation network.
    
    \begin{figure*}[!ht]
\vskip -0.1in
    \begin{center}
        \centerline{\includegraphics[width=2\columnwidth]{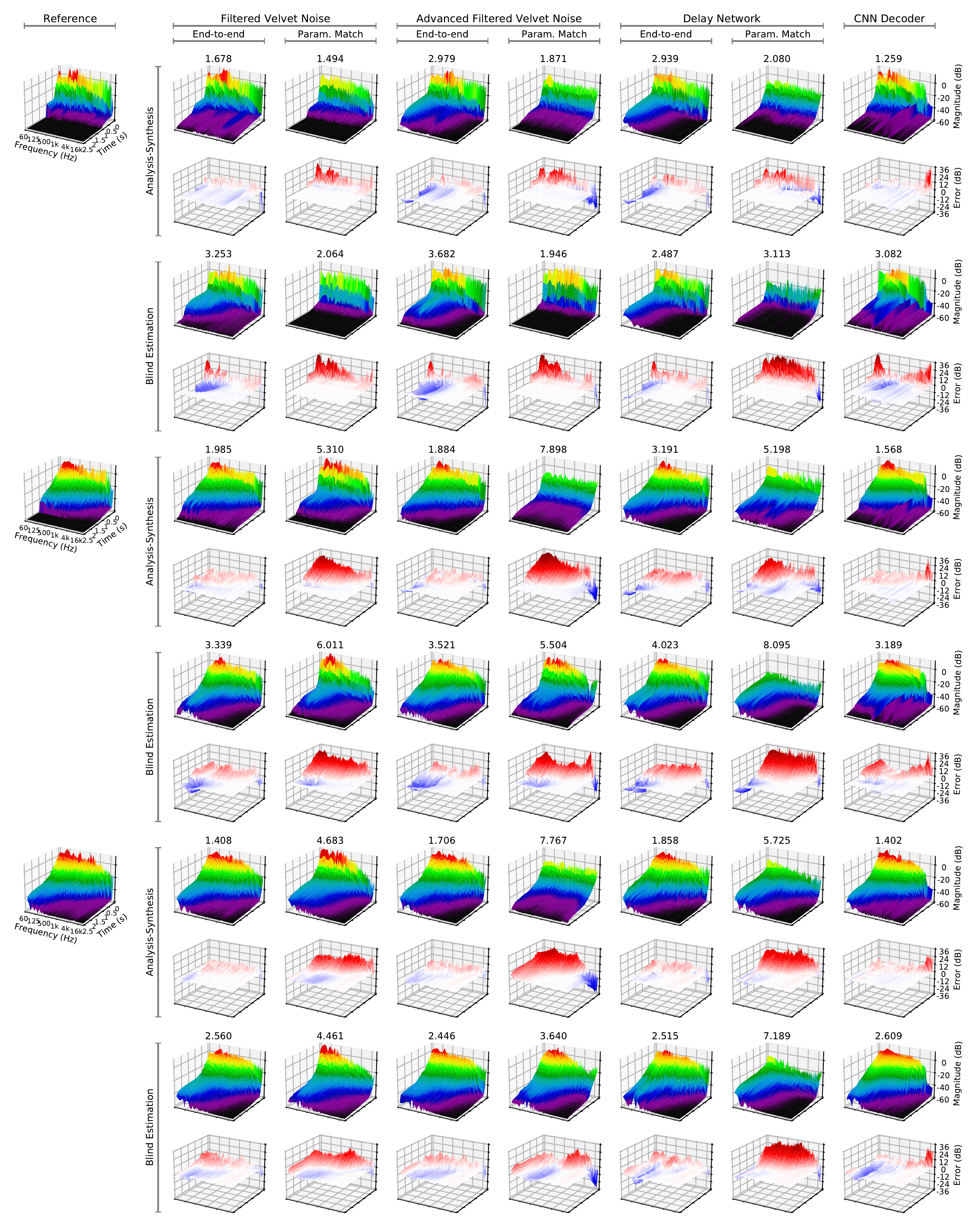}}
        \caption{EDR plots of the reference RIRs and their estimations with the trained networks. For each reference, a total of twelve networks' estimations and their errors are visualized where each network is trained for one of the three AR models (FVN, AFVN, and DN), one of the two tasks (analysis-synthesis and blind estimation), and one of the two training approaches (end-to-end learning with the proposed DAR models and the baseline non-end-to-end parameter-matching learning). The number provided above each plot is the EDR distance of the estimation. }
        \label{fig:edr}
    \end{center}
\end{figure*}

    \subsection{AR Model Efficiency and Estimation Performance}
        Each AR model approximates {reference} reverberation differently, and consequently, its computational efficiency and its estimation networks' performance varies.
        
        \subsubsection{Filtered Velvet Noise} $\text{FVN}$ filters each source segment independently (see Figure \ref{fig:magnitude-response}). Such flexibility enables its parameter estimators to perform better than the other AR models' counterparts experimented in this paper. However, at the same time, it also has the largest number of ARPs ($930$), which makes it cumbersome to control manually. Also, it could be computationally expensive to use multiple $\text{FVN}$ instances in real-time ($2546$ FLOPs for each).
        
        \subsubsection{Advanced Filtered Velvet Noise}
        Instead, $\text{AVFN}$ optimizes each coloration filter by decomposing it into the initial coloration and delta-coloration filters, making the model more controllable and efficient ($360$ ARPs and $1511$ FLOPs) than the $\text{FVN}$. Such simplification costs estimation performance since the underlying assumption of the optimization that coloration of real-world RIRs changes gradually does not necessarily hold.
        In addition, the AFVN networks could be more challenging to optimize than the FVN's since each segment's coloration depends on the previous ones'. Indeed, we empirically observed that their training losses decreased slower than the FVN's. 
        
        \subsubsection{Delay Network}
        In DN, we simplified the filter structure even further by using a single absorption filter for each feedback loop. Similar to the $\text{AVFN}$ case, $\text{DN}$ trade-offs its estimators' performance with its controllability and efficiency ($200$ ARPs and $1285$ FLOPs). Again, the estimation performance degradation comes from the $\text{DN}$'s restricted expressibility (forced exponential decay) and training difficulty.

\section{Conclusion} \label{section:conclusion}
    We proposed differentiable artificial reverberation (DAR) models that can be integrated with deep neural networks (DNNs). Among numerous pre-existing artificial reverberation (AR) models, we selectively implemented Filtered Velvet Noise (FVN), Advanced Filtered Velvet Noise (AFVN), and Delay Network (DN) differentiably. Nonetheless, the proposed method, which replaces the infinite impulse response (IIR) components with finite impulse response (FIR) approximations via frequency-sampling, is applicable to any other AR model. Then, using the DAR models, we trained the proposed artificial reverberation parameter (ARP) estimation networks end-to-end. The evaluation results showed that our networks captured the target reverberation accurately in both analysis-synthesis and blind estimation tasks. We showed, in particular, that the end-to-end training significantly improves the estimation performance at the tolerable cost of the approximation errors caused by the frequency-sampling. Additionally, we demonstrated that the structural priors of the AR models avoid perceptual artifacts that a DNN as a room impulse response (RIR) generator might produce. Thus, our framework successfully combined powerful and fully general deep learning techniques with well-established domain knowledge of reverberation.
    
    Here, we outline the remaining challenges and future work. First, we used the shoe-box simulation to obtain a large amount of training data, which is slightly different from the real-world RIRs.
    Interestingly, energy decay relief (EDR) errors of the end-to-end models showed similar patterns regardless of the equipped AR/DAR model (see Figure \ref{fig:edr}). This suggests that the data characteristics discrepancy might be the cause of the performance degradation. Therefore, collecting more real-world RIRs and using powerful augmentation methods could replace the simulation-based data and improve the performance. Second, we only used the RIRs and the reverberant speech signals as references. However, other possible applications with different reference types exist (for example, automatic mixing of musical signals). 
    Finally, we investigated how each AR model can be compared in terms of its expressive power and difficulty of parameter optimization under the deep learning environment. The evaluation results revealed a trade-off relationship; when a more compact AR model is used, the estimation performance was upper-bounded by its limited expressive power. Also, the proposed modifications to the original AR models improved estimation performance and training speed. 
    All of these indicate that other than the presented AR models and estimation networks, there may be more ``deep-learning-friendly'' models and ``AR-model-friendly'' DNNs that yield better performance-efficiency trade-offs, and seeking those is left as future work.
\appendices
    \section{Ablations and Comparisons of the AR/DAR Models} \label{appendix:ablations} 
    \subsection{AR/DAR Model Ablations} \label{subsection:ablations}
    \begin{table}[t]
\caption{Ablation on the AR/DAR Model configurations.}
\label{table:ar-ablation}
\setlength\tabcolsep{3.2pt}
\begin{center}
\begin{footnotesize}
\begin{sc}
\begin{tabular}{c|c|cc|cc|cc}
\toprule
\multirow{2}{*}{Model} & $\mathcal{L}_\text{Match}$ & \multicolumn{2}{c|}{$\Delta\text{T}_{30} \, (\%)$}  & \multicolumn{2}{c|}{$\Delta\text{DRR} \, (\si{dB})$} & \multicolumn{2}{c}{$\Delta\text{C}_{50} \, (\si{dB})$} \\
& {\scalebox{0.8}{$(\times 10^{-1})$}} & Full & Freq & Full & Freq & Full & Freq \\
\midrule
FVN & $\mathbf{1.236}$ & $\mathbf{1.89}$ & $\mathbf{4.20}$ & $\mathbf{0.94}$ & $\mathbf{3.39}$ & $0.60$ & $\mathbf{1.95}$ \\
$T_i=10$ & $1.257$ & $2.74$ & $6.14$ & $1.56$ & $4.21$ & $\mathbf{0.48}$ & $2.25$ \\
$\Delta l_i = L/S$ & $1.255$ & $2.56$ & $5.43$ & $1.48$ & $3.68$ & $0.89$ & $2.29$\\
$-h_0[n]$ & $1.250$ & $2.23$ & $10.71$ & $3.17$ & $11.94$ & $0.81$ & $9.88$\\
$M=1$ & $1.261$ & $2.53$ & $4.42$ & $1.51$ & $3.65$ & $0.82$ & $2.21$\\
\midrule 
AFVN & $\mathbf{1.277}$ & $\mathbf{4.77}$ & $\mathbf{9.96}$ & $1.59$ & $\mathbf{3.64}$ & $\mathbf{0.61}$ & $\mathbf{2.06}$ \\
$-C_1$ & $1.446$ & $6.75$ & $12.60$ & $\mathbf{1.46}$ & $8.43$ & $1.76$ & $3.64$\\
$-\Delta C_j$ & $1.324$ & $7.12$ & $32.16$ & $2.99$ & $5.01$ & $0.92$ & $2.62$\\
$-C_1, \Delta C_j$ & $1.695$ & $8.91$ & $28.21$ & $2.15$ & $6.62$ & $2.64$ & $5.89$ \\
\midrule 
DN & $1.334$ & $8.47$ & $11.59$ & $1.71$ & $3.83$ & $1.08$ & $2.30$ \\
$-C_1$ & $1.446$ & $10.47$ & $20.75$ & $3.69$ & $4.55$ & $2.31$ & $3.03$\\
$-\mathbf{C}_\Delta$ & $1.351$ & $7.83$ & $31.22$ & $2.22$ & $4.71$ & $0.94$ & $2.38$\\
$-C_1, \mathbf{C}_\Delta$ & $1.729$ & $11.62$ & $31.20$ & $4.00$ & $6.68$ & $2.03$ & $4.54$  \\
$-\mathbf{Q}_n$ & $1.307$ & $5.93$ & $12.48$ & $1.81$ & $3.93$ & $1.08$ & $2.39$\\
$-\mathbf{U}$ & $1.535$ & $21.10$ & $\mathbf{10.76}$ & $1.03$ & $3.44$ & $0.88$ & $2.28$\\
$-\mathbf{U}, \mathbf{Q}_n$ & $\mathbf{1.251}$ & $\mathbf{7.43}$ & $10.79$ & $\mathbf{1.00}$ & $\mathbf{3.39}$ & $\mathbf{0.70}$ & $\mathbf{2.09}$\\
\bottomrule
\end{tabular}

\end{sc}
\end{footnotesize}
\end{center}
\end{table} 
    \begin{table}[!t]
\caption{Differentiable FVN with various coloration filters and parameterization approaches.}
\label{table:filter}
\setlength\tabcolsep{3.2pt}
\begin{center}
\begin{footnotesize}
\begin{sc}
\begin{tabular}{c|c|c c|c c|c c}

\toprule
\multirow{2}{*}{Filter} & $\mathcal{L}_\text{Match}$ & \multicolumn{2}{c|}{$\Delta\text{T}_{30} \, (\%)$}  & \multicolumn{2}{c|}{$\Delta\text{DRR} \, (\si{dB})$} & \multicolumn{2}{c}{$\Delta\text{C}_{50} \, (\si{dB})$} \\

& {\scalebox{0.8}{$(\times 10^{-1})$}} & Full & Freq & Full & Freq & Full & Freq \\
\midrule 
SSVF & $1.236$ & $\mathbf{1.89}$ & $\mathbf{4.20}$ & $\mathbf{0.94}$ & $\mathbf{3.39}$ & $0.60$ & $1.95$\\
PSVF & $\mathbf{1.235}$ & $1.95$ & $4.52$ & $0.96$ & $3.64$ & $0.53$ & $\mathbf{1.86}$ \\
PEQ & $1.254$ & $2.50$ & $5.47$ & $0.97$ & $3.61$ & $0.62$ & $1.89$ \\
SBIQ & $1.322$ & $3.83$ & $9.44$ & $1.55$ & $4.61$ & $0.59$ & $2.38$\\
PBIQ & $1.337$ & $4.88$ & $11.77$ & $1.57$ & $4.01$ & $0.62$ & $2.19$\\
FIR & $1.370$ & $2.72$ & $10.10$ & $1.80$ & $3.73$ & $\mathbf{0.51}$ & $2.16$\\
LFIR & $1.384$ & $3.54$ & $11.84$ & $1.64$ & $4.31$ & $0.62$ & $2.65$\\
\bottomrule
\end{tabular}

\end{sc}
\end{footnotesize}
\end{center}
\end{table}

        We conducted ablation studies to verify that the modifications to the original AR models improve the estimation performance. We evaluated the modified models with the analysis-synthesis task since their performance differences remained the same regardless of the task in our initial experiments. 

        Table \ref{table:ar-ablation} summarizes the results. 
        FVN, {AFVN}, and DN denote the proposed models. 
        $T_i=10$, $\Delta l_i = L/S$, $-h_0[n]$, and $M = 1$ denote the proposed FVN model without non-uniform pulse distance and allpass filters, non-uniform segmentation, a deterministic FIR, and finer segment gains. Their results show that the proposed modifications contribute to better performance.
        We omitted the same ablations for the {AFVN}. Instead, we discarded the initial filter $C_1$, delta filter $\Delta C_j$, or both. We also evaluated the DN without the initial filter $C_1$ or delta filter $\mathbf{C}_\Delta$. The results reveal the importance of modelling the frequency-dependent nature of reverberation. 

        For DN, $-\mathbf{Q}_n$, $-\mathbf{U}$, and $-\mathbf{U}, \mathbf{Q}_n$ denote DN without the time-varying mixing matrix $\mathbf{Q}_n$, the allpass filters $\mathbf{U}$, and both, respectively. While $-\mathbf{U}, \mathbf{Q}_n$ reports best evaluation metrics, its low echo density produced unrealistic sound. Using both $\mathbf{U}$ and $\mathbf{Q}_n$ improved plausibility of reverberation with an acceptable performance degradation. Refer to the provided audio samples.

    \subsection{Comparisons on Different Coloration Filters} \label{subsection:coloration-comparison}
        We compared different IIR/FIR filters and parameterization approaches. We evaluated them with the analysis-synthesis task using FVN. Table \ref{table:filter} summarizes the results. 
        SSVF, PSVF, PEQ, SBIQ, PBIQ, FIR, and LFIR denote serial SVF, parallel SVF, PEQ, serial biquad, parallel biquad, FIR, and linear-phase FIR from \cite{engel2020ddsp}. 
        The order of every filter was set to $40$. The activation functions and bias initialization for each filter's decoder are as follows. PSVF has the same setting as SSVF. For SBIQ and PBIQ decoders, we used activations from \cite{nercessian2021lightweight}, and their biases are initialized to match that of SSVF and PSVF. For PEQ, unlike DN, we used $10^{x}$ for the activation of the gain $\mathbf{G}$. FIR/LFIR have no activation and custom bias initialization.

        In spite of having identical expressive power, SSVF/PSVF performed better than SBIQ/PBIQ by large margins. PEQ performed slightly worse than SSVF due to its restricted degree of freedom. FIR/LFIR also performs worse than SSVF/PSVF and PEQ because they cannot change their frequency responses radically as other IIR filters could.
        
    \begin{table}[!t]
\begin{threeparttable}
\caption{
Effects of frequency-sampling resolution and regularization on the match loss $\mathcal{L}_\text{Match} \,(\times 10^{-1})$.}
\setlength\tabcolsep{2.8pt}
\label{table:sampling-points}
\begin{center}
\begin{footnotesize}
\begin{sc}
\begin{tabular}{c|cc|cc|cc|cc}
\toprule
\multirow{3}[4]{*}{Model} & \multicolumn{8}{c}{Frequency-Sampling Resolution} \\
\cmidrule{2-9}
& \multicolumn{2}{c|}{$\times 1$ (Full)} & \multicolumn{2}{c|}{$\times 0.5$} & \multicolumn{2}{c|}{$\times 0.25$} & \multicolumn{2}{c}{$\times 0.125$} \\
\cmidrule{2-9}
& DAR & AR & DAR & AR & DAR & AR & DAR & AR \\
\midrule
$\text{FVN}_{\beta = 1}$ & $\mathbf{{1.236}}$&$\mathbf{{1.236}}$ & ${1.259}$&$1.259$ & ${1.252}$&$1.249$ & ${1.574}$&$1.570$\\
$\text{FVN}_{\beta = 0}$ & $\mathbf{{{1.231}}}$&${1.237}$ & ${1.283}$&$1.282$ & ${1.264}$&$1.263$ & ${1.280}$&$1.773$\\
DN & $\mathbf{{1.277}}$&${1.307}$ & $-$& $1.328$ & $-$& $1.361$ & $-$& $1.393$ \\
\bottomrule
\end{tabular}
\end{sc}
\end{footnotesize}
\end{center}

\end{threeparttable}
\end{table}

    \subsection{Effects of the Frequency-Sampling Resolution} \label{subsection:reliability}
        Table \ref{table:sampling-points} demonstrates the effects of the frequency-sampling resolution and the regularization loss $\mathcal{L}_{\text{\begin{sc}Reg\end{sc}}}$.
        From the table, $\times x$ denotes the relative sampling resolution compared to the default configuration, e.g., $\times 0.5$ denotes $N=2\si{k}$ for the FVN and $N=6\si{k}$ for the DN. For each resolution and regularization setup, we report match loss with the DAR and AR models. Results of the differentiable DNs with lower resolutions are omitted since they are calculated with shorter IRs.
        
        As expected, higher sampling resolution led to smaller loss difference. Also, introducing the regularization term ($\beta=1$) reduced the difference further. As a result, the final differentiable FVN model (with full resolution and regularization) showed little loss difference.
        Moreover, higher resolution led to better performance, which might be because increased resolution helped each network to find better ARPs. 
        \section{Details on the Baseline Methods}  \label{appendix:parameter-match-baseline}
        
    \subsection{ARP Match Models}
        \subsubsection{Training}
        We trained the ARP match baseline models as follows. First, we randomized ARPs $\mathbf{P}_1, \cdots, \mathbf{P}_n$ and generated an IR with using the DAR model. Then, each baseline network estimated ARPs $\hat{\mathbf{P}}_1, \cdots, \hat{\mathbf{P}}_n$ from the given IR. We evaluated the estimation and trained the baseline using an ARP-match loss $\mathcal{L}_\text{ARP}$ defined as follows,
        \begin{equation}
            \mathcal{L}_\text{ARP}(\mathbf{P}_i, \hat{\mathbf{P}}_i) = \sum_i \alpha_i \left\| f_i(\mathbf{P}_i) - f_i(\hat{\mathbf{P}}_i) \right\|_1.
        \end{equation} 
        Here, the index $i$ denotes a different ARP tensor, e.g., $\mathbf{P}_0 = \mathbf{g}$ and $\mathbf{P}_1 = \mathbf{h}_0$. $f_i(\cdot)$ and $\alpha_i$ are an elementwise function and a constant, respectively. We set $\alpha_i=10$ for the all DAR models' bypass FIRs and the DN's absorption filter parameters. We used $f_i(x)=\log_{10}(x)$ for the FVN and {AFVN} segment gains. 
        We chose $f_i(x) = x$ and $\alpha_i=1$ otherwise.

    \subsubsection{Network Architecture}
        We trained an almost identical network to the proposed network for each model/task. The only difference is that we added an activation $10^{-\zeta(x)/20}$ to the $\mathbf{g}$ decoders of FVN and {AFVN} to improve their performance. 

    \subsubsection{Data Generation} 
        We tuned each DAR model's ARP randomization scheme to match the synthesized IRs' reverberation parameter statistics to the validation set. For the FVN baselines, we first sampled a reverberation time $\text{T}_\text{30} \sim \mathcal{U}_{\log}[50\si{ms}, 8\si{s})$ then generated the segment gains $\mathbf{g}$ that match the $\text{T}_\text{30}$. Here, $\mathcal{U}_{\log}[\cdot, \cdot)$ denotes a uniform distribution in log scale. In this procedure, we also compensated the average pulse distance by dividing each gain $g_i$ by $\sqrt{T_i}$. For the initial coloration filter $C_1$, we first generated PEQ parameters as $\omega_i \sim \mathcal{U}_{\log}[40\si{Hz}, 16\si{kHz})$, $R_i \sim \mathcal{U}_{\log}[0.2, 5)$, and $G_i \sim \mathcal{U}[-18\si{dB}, 18\si{dB})$ (we sorted the cutoff frequencies after the sampling) and derived their SVF parameters. After that, we perturbed each parameter's value slightly. This ARP generation method was motivated by the observation that the differentiable SVFs act like a relaxed PEQ (see Figure \ref{fig:magnitude-response}). We gradually changed the parameters of $C_1$ to obtain $C_2, \cdots, C_K$, modeling the frequency-dependent decay. Finally, we set $h_0[n]$ as an uniform noise with a gain sampled from $\mathcal{U}[-24\si{dB}, 0\si{dB})$. 
        We followed similar procedures for the AFVN baselines. One difference is that the first delta filter $C_{\Delta 2}$ is randomized and the delta filters afterwords are slight deviation of the first one. For the DN baselines, we sampled each absorption PEQ's gain with $G_i \sim \mathcal{U}[-2.8\si{dB}, 0\si{dB})$.
    
    \subsection{DNN Decoder Models}
        Regarding the first and second axis of the shared latent $\mathbf{z}$ as a time and channel axis, respectively, our decoder upsamples $\mathbf{z}$  with seven one-dimensional transposed convolution layers. Configurations of these layers are as follows. Channels: $128$, $128$, $64$, $64$, $64$, $64$, then $64$. Kernel sizes: $5$, $5$, $5$, $5$, $5$, $5$, then $3$. Strides: $3$, $3$, $3$, $3$, $3$, $3$, then $2$. Furthermore, after each upsampling, we inserted a small temporal convolutional network with configurations as follows. Channels: $128$, $128$, $64$, $64$, $64$, $64$, and $64$. The number of layers: $3$, $3$, $2$, $2$, $2$, $2$, and $2$. Kernel sizes: all $7$. Finally, we added $1\times 1$ convolution as the last layer to mix all channels. We used the ReLU activation between the layers. This decoder has about $2.5\si{M}$ parameters. 

\section{Details on the Frequency-sampling Method} \label{appendix:freq-samp}
    \subsection{Proof of Equation \ref{eq:fr-deviation}}
        With equation \eqref{eq:time-alias}, the time-aliasing error can be written as 
        \begin{equation}
            \|H-H_N\|_2^2 = \sum_{n=0}^{N-1}\left|\sum_{m=1}^{\infty}h[mN+n]\right|^2 +  \sum_{n=N}^{\infty} \left| h[n] \right|^2. 
            \label{eq:l2-time-domain-deviation}
        \end{equation}
        With the triangle inequality and $\|X\|_2^2\leq \|X\|_1^2$, we can upper-bound $\|H-H_N\|_2^2$ with $2S_N^2$ where $S_N = \sum_{n=N}^{\infty} \left| h[n] \right|$ is an absolute sum of the tail of the original IR $h[n]$ $(n\geq N)$. 
        Next, we expand $h[n]$ with partial fraction expansion \cite{liski19parallel, FILTERS07}. For $H(z)$ with $M$ distinct poles where each of them $\nu_i \in \mathbb{C}$ has multiplicity $r_i \in \mathbb{N}$, we can express $h[n]$ as a sum of an FIR $h^\text{FIR}[n]$ and IIRs $h^\text{IIR}_{i,k}[n]$ where following holds,
        \begin{equation}
            H(z) = H^\text{FIR}(z) + \sum_{i=1}^{M} \sum_{k=1}^{r_i} \underbrace{\frac{\zeta_{i,k}}{(1-\nu_iz^{-1})^k}}_{H^\text{IIR}_{i,k}(z)}. \quad \left(\zeta_{i,k} \in \mathbb{C}\right)
        \end{equation}
        Then, applying the triangle inequality to the summation gives an upper bound $U_N$, an absolute sum of the IIRs' tails. 
        \begin{equation}
            S_N \leq  
            \sum_{i=1}^{M} \sum_{k=1}^{r_i} \sum_{n=N}^{\infty} \left| h^\text{IIR}_{i, k}[n] \right| = U_N. \label{eq:pfe-inequality}
        \end{equation}
        From above equation, each IIR, first sum, and second sum show  $O(n^{k-1}|\nu_i|^{n})$, $O(N^{k-1}|\nu_i|^{N})$, and $O(N^{r_i-1}|\nu_i|^{N})$ asymptotic behavior, respectively \cite{FILTERS07}.
        We conclude the proof: 
        \begingroup
        \setlength{\thinmuskip}{1mu}
        \setlength{\medmuskip}{2mu}
        \setlength{\thickmuskip}{3mu}
        \begin{equation}
            \|H-H_N\|_2 \leq \sqrt{2}S_N \leq \sqrt{2}U_N = \sum_{i=1}^{M} O(N^{r_i-1}\left|\nu_i\right|^{N}).
        \end{equation}
        \endgroup

    \subsection{Proof of Equation \ref{eq:deviation-loss}}
        We denote $H(e^{j\omega})$ and $\partial X / \partial p$ with $H$ and $X'$. $X_N'$ denotes $(X_N)'=(X')_N$.
        We upper-bound the gradient difference $\Delta \mathcal{G}$ using integral inequalities as
        \begingroup
        \setlength{\thinmuskip}{1mu}
        \setlength{\medmuskip}{2mu}
        \setlength{\thickmuskip}{3mu}
        \begin{equation}
            \Delta \mathcal{G} \leq \underbrace{\int_0^{2\pi} | H_\text{Ref}(H' - H_N') |\frac{d\omega}{\pi}}_{\leq 2\|H_\text{Ref}\|_2\|H- H_N'\|_2} + 
            \underbrace{\int_0^{2\pi} |HH' - H_NH_N'|\frac{d\omega}{\pi}}_{\leq 2\|HH'- H_NH'_N\|_2} 
        \end{equation}
        \endgroup
        where the under-braced ones are the Cauchy-Schwartz inequalities.
        Considering the first integral's upper bound, $\|H_\text{Ref}\|_2$ is a finite constant. In $\|H'-H_N'\|_2$, $H'$ is another stable LTI filter if $p$ is well-defined.
        If $\nu_i$ is a function of $p$, its multiplicity $r_i$ is doubled in $H'$. In the worst case, $p$ controls every pole of $H$ and $\|H'-H_N'\|_2=\sum_{i=1}^{M} O(N^{2r_i-1} |\nu_i|^{N})$ holds.    
        Similarly, the upper bound of the second integral consists an $l_2$ distance between $HH'$ and $H_NH_N'$. Since $HH'$ has $2r_i$ or $3r_i$ multiplicity for each pole $\nu_i$, $\|HH'-H_NH'_N\|_2=\sum_{i=1}^{M} O(N^{3r_i-1} |\nu_i|^{N})$ holds at the worst case, which concludes the proof.
\ifCLASSOPTIONcaptionsoff
  \newpage
\fi
\bibliography{refs}
\bibliographystyle{IEEEtran}

\end{document}